\documentclass[12pt]{article}

\usepackage{jheppub,amsbsy,amssymb,latexsym,amsfonts,amsmath,amscd,epsfig,multicol,bbm}

\newcommand{\alg}[1]{\mathfrak{#1}}
\newcommand{\gen}[1]{\mathrm{#1}}
\newcommand{\wt}[1]{\widetilde{#1}}

\newcommand{\cA}{{\cal A}}
\newcommand{\cI}{{\cal I}}
\newcommand{\cJ}{{\cal J}}
\newcommand{\cL}{{\cal L}}
\newcommand{\cF}{{\cal F}}
\newcommand{\cG}{{\cal G}}
\newcommand{\cR}{{\cal R}}
\newcommand{\cU}{{\cal U}}
\newcommand{\cY}{{\cal Y}}

\newcommand{\ga}{\gamma}
\newcommand{\de}{\delta}
\newcommand{\ep}{\epsilon}
\newcommand{\ve}{\varepsilon}
\newcommand{\la}{\lambda}

\newcommand{\Tr}{{\rm Tr}} 
\newcommand{\pexp}{{\rm P}\!\exp} 
\newcommand{\sch}{Schr\"odinger } 
\newcommand{\copro}{\Delta } 
\newcommand{\coproop}{\widetilde{\Delta} } 
\newcommand{\no}{\nonumber}
\newcommand{\e}{{\rm e}}

\allowdisplaybreaks

\title{Schr\"odinger Sigma Models and Jordanian Twists}

\author{Io Kawaguchi$^{\ast}$\footnote{E-mail:~io@gauge.scphys.kyoto-u.ac.jp}, 
Takuya Matsumoto$^{\dagger}$\footnote{E-mail:~tmatsumoto@usyd.edu.au} 
and Kentaroh Yoshida$^{\ast}$\footnote{E-mail:~kyoshida@gauge.scphys.kyoto-u.ac.jp}}
\affiliation{$^{\ast}${\it Department of Physics, Kyoto University Kyoto 606-8502, Japan} \\
$^{\dagger}${\it School of Mathematics and Statistics, University of Sydney, NSW 2006, Australia} }

\arxivnumber{1305.6556[hep-th]}

\abstract{
We proceed to study the integrable structures of two-dimensional 
non-linear sigma models defined on three-dimensional Schr\"odinger spacetimes. 
We show that anisotropic Lax pairs are equivalent with 
isotropic Lax pairs with flat conserved currents under non-local gauge transformations. 
Then a quite non-trivial realization of the undeformed Yangian symmetry 
$\mathcal{Y}(\mathfrak{sl}(2))$ is revealed after an appropriate gauge fixing, which is determined by 
comparing the gauge transformation to a quantum Jordanian twist. 
As a result, an exotic symmetry found in 
\href{http://arxiv.org/abs/1209.4147}{[arXiv:1209.4147]} 
may be interpreted as a Jordanian deformation of $\mathcal{Y}(\mathfrak{sl}(2))$. 
}


\begin{document}

\maketitle

\section{Introduction}

One of the most striking achievements in the study of the AdS/CFT correspondence \cite{M,GKP,W} 
is the discovery of the integrable structure (For a comprehensive review, see \cite{review}). 
On the string-theory side, it is closely related to the classical integrability of two-dimensional 
non-linear sigma models on symmetric cosets \cite{BPR}. The symmetric cosets which can appear 
as consistent string backgrounds are classified in \cite{Zarembo-symmetric}, including supercharges.  

\medskip 

The next step is to consider integrable deformations of AdS spaces and spheres 
(For the works related to our issue, for example, see \cite{Cherednik,FR,BFP,ORU,BR}). 
The deformed geometries are still represented by cosets \cite{SYY} 
and are homogeneous, while the symmetric-space structure is lost. Thus the familiar way 
in symmetric spaces \cite{Luscher1,Luscher2,BIZZ,Bernard-Yangian,MacKay,AAR} is not applicable 
to discuss the integrable structure in the deformed cases any more. 
However, some new aspects appear due to the deformation and 
the integrable structure is enriched as in analogy with 
the deformations from the XXX models to the XXZ models. 

\medskip 

For example, non-linear sigma models defined on squashed S$^3$ have a rich structure. 
The squashing breaks the isometry of round S$^3$\,, 
$SO(4) =SU(2)_{\rm L} \times SU(2)_{\rm R}$ to $SU(2)_{\rm L} \times U(1)_{\rm R}$\,. 
The sigma models have a couple of Yangians based on $SU(2)_{\rm L}$ \cite{KY}. 
It is shown that the Yangians are preserved even after adding 
the Wess-Zumino term \cite{KOY}. One may also consider an infinite-dimensional symmetry 
originated from $U(1)_{\rm R}$\,. This is a quantum affine algebra at the classical level \cite{KYhybrid,KY-summary,KMY-QAA}. 
In the end, two different kinds of infinite-dimensional symmetries are realized simultaneously 
and hence the integrable structure should be called the 
``hybrid integrability''. The two algebras are gauge-equivalent 
\cite{KMY-monodromy}. 

\medskip 

It is also interesting to consider the classical integrable structure of the sigma models 
defined on three-dimensional Schr\"odinger spacetimes. The isometry is given by 
$SL(2,{\mathbb R})_{\rm L} \times U(1)_{\rm R}$\,.  
It has been shown that the sigma models possess a pair of $SL(2,{\mathbb R})_{\rm L}$ Yangians and an affine 
extension of $q$-deformed Poincar\'e algebra (named an exotic symmetry) \cite{KY-Sch,exotic}. 
The latter symmetry is based on $U(1)_{\rm R}$\,. The mathematical background of the affine extension has not been elucidated. 

\medskip 

In this paper we proceed to study the affine extension of $q$-deformed Poincar\'e algebra. 
Our aim is to unveil the mathematical background. 
We first construct a conserved current satisfying the flatness condition 
by performing a gauge transformation for the monodromy matrix based on $U(1)_{\rm R}$\,. 
Then the BIZZ construction \cite{BIZZ} leads to  $SL(2,{\mathbb R})_{\rm R}$ 
Yangian in a quite non-trivial representation. As a result, the gauge transformation can be identified as 
undoing the classical Jordanian twist. That is, 
the exotic symmetry found in \cite{exotic} may be interpreted as the Jordanian twist of the Yangian. 

\medskip 

This paper is organized as follows. In Section \ref{sec:setup}, we introduce the classical action of 
the Schr\"odinger sigma models and give a short review of the monodromy matrix based on $U(1)_{\rm R}$\,. 
In Section \ref{sec:Jordan},  we construct a conserved current satisfying the flatness condition 
by performing a gauge transformation to the monodromy matrix. 
Then the BIZZ construction leads to $SL(2,{\mathbb R})_{\rm R}$ 
Yangian in a quite non-trivial representation. 
In Section \ref{sec:Jordanian}, the mathematical formulation of quantum Jordanian twist is introduced. 
The relation to the gauge-fixing condition is explained from the point of view of the Jordanian twist.  
In Section \ref{sec:left}, we argue the relation of the Jordanian twist 
to the Yangian generators based on $SL(2,{\mathbb R})_{\rm L}$ through the left-right duality. 
In Section \ref{sec:dipole}, we discuss the geometric interpretation of 
the classical Jordanian twist. 
Section \ref{sec:concl} is devoted to conclusion and discussion. 

\medskip 

The Appendix part might seem lengthy because messy computations, for example, of current algebras are explained in detail. 
We, however, believe that the calculation process would be valuable for the readers who are interested in the detailed computations.  
In Appendix \ref{app:twistF}, we will show the detailed computations in performing a gauge transformation to the monodromy matrix 
and rewriting the conserved currents. 
Appendix \ref{ca:app} is devoted to illustrate the computation of the current algebra.  
In Appendix \ref{app:gfix}, as a side note, we try to argue another flat conserved current, 
for the twist operation does {\it not} correspond to the Jordanian twist. 
Then the resulting algebra is a deformed Yangian which contains the third order terms even at the level-zero.

\section{Setup \label{sec:setup} }

Let us first introduce the classical action of two-dimensional non-linear sigma models defined 
on three-dimensional Schr\"odinger spacetimes. Then we give a brief review of the Lax pairs and 
the monodromy matrices based on $U(1)_{\rm R}$\,. The behavior of them under gauge transformations 
is explained. Finally we give a review on an infinite dimensional extension 
of the $q$-deformed Poincar\'e algebra, which is referred to as the exotic symmetry.

\subsection{Schr\"odinger spacetimes }

Schr\"odinger spacetimes in three dimensions are known as null-like deformations of AdS$_3$\,. 
The metric is given by \cite{Israel:2004vv,Son,BM}
\begin{align}
ds^2=L^2\left[-2{\rm e}^{-2\rho}dudv+d\rho^2-C{\rm e}^{-4\rho}dv^2\right]\,. 
\label{Sch3-angle}
\end{align}
The deformation is measured by a real constant parameter $C$\,. When $C=0$\,, the metric (\ref{Sch3-angle}) 
describes the AdS$_3$ space with the curvature radius $L$ and the isometry is 
$SO(2,2)=SL(2,{\mathbb R})_{\rm L}\times SL(2,{\mathbb R})_{\rm R}$\,.  
When $C\neq 0$\,, $SL(2,{\mathbb R})_{\rm R}$ is broken to $U(1)_{\rm R}$ while 
$SL(2,{\mathbb R})_{\rm L}$ is preserved. That is, the global symmetry that survives the deformation is 
$SL(2,{\mathbb R})_{\rm L}\times U(1)_{\rm R}$ in total. 
This symmetry yields two distinct pictures to describe the classical dynamics, as we will see later.  

\medskip 

It is convenient to rewrite the metric \eqref{Sch3-angle} 
in terms of the left-invariant one-form,
\begin{align}
J &= g^{-1}dg  & 
g&={\rm e}^{2vT^+}{\rm e}^{2\rho T^2}{\rm e}^{2uT^-} \in SL(2,{\mathbb R}) \no \\ 
&=-T^+J^--T^-J^++T^2J^2 &J^a&=2{\rm Tr}(T^aJ) \,. 
\end{align}
The $\alg{sl}(2)$ generators $T^a$ ($a=\pm,2$) satisfy the commutation relations 
$[T^a,T^b ]=\varepsilon^{ab}{}_cT^c$\,, where the structure constants $\ve^{ab}{}_c$ 
are the totally anti-symmetric tensor normalized as $\ve^{-+}{}_2=+1$\,. 
The generators are normalized as $\Tr (T^aT^b)=\tfrac{1}{2}\ga^{ab}$ with the Killing metric $\gamma^{ab}$\,. 
The $\alg{sl}(2)$ indices are raised and lowered by $\gamma^{ab}$ and its inverse $\ga_{ab}$ respectively. 
From here on, we work with the fundamental representation, 
\begin{align}
&T^+=\frac{1}{\sqrt{2}} \biggl( \; \begin{matrix} 0 & ~~1 \\ 0 & ~~0 \end{matrix}\; \biggl)\,, \quad
T^-=\frac{-1}{\sqrt{2}} \biggl( \; \begin{matrix} 0 & ~~0 \\ 1 & ~~0 \end{matrix}\; \biggl)\,, \quad 
T^2=\frac{1}{2} \biggl( \; \begin{matrix} 1 & 0 \\ 0 & -1 \end{matrix}\; \biggl)\,. 
\end{align}
Note that $T^2$ is taken as the Cartan generator.   

\medskip 

By using this one-form $J$,  the metric \eqref{Sch3-angle} can be rewritten into the following form, 
\begin{align}
ds^2=\frac{L^2}{2} \bigl[\Tr(J^2)-2C\Tr(T^-J)^2 \bigr]\,.  
\label{last}
\end{align}
This expression enables us to regard the \sch spacetimes as null-deformations of AdS$_3$. 
Now it is manifest that the metric \eqref{last} is invariant under the 
$SL(2,{\mathbb R})_{\rm L}\times U(1)_{\rm R}$ transformations defined as 
\begin{eqnarray}
g\mapsto {\rm e}^{\beta_a T^a} \, g \,{\rm e}^{-\alpha T^-}\,, \qquad 
J\mapsto {\rm e}^{\alpha T^-} \, J \, {\rm e}^{-\alpha T^-}  
\end{eqnarray}
with any $\beta_a~(a=\pm,2),~\alpha\in\mathbb{R}$\,.

\subsection{\sch sigma models}

Let us introduce here the classical action of two-dimensional non-linear sigma models defined 
on \sch  spacetimes in three dimensions. For simplicity, we shall refer to those as {\it \sch sigma models}. 

\medskip 

With the metric \eqref{last}, the classical action and the Lagrangian is given by 
\begin{align}
S=\int^\infty_{-\infty}\!\!\!dt\!\int^\infty_{-\infty}\!\!\!dx\, L[J] 
\quad \text{with} \quad 
L[J]=-\eta^{\mu\nu} \bigl[\Tr(J_{\mu}J_{\nu})-2C\Tr(T^{-}J_{\mu})\Tr(T^{-}J_{\nu})\bigr]\,. 
\label{action}
\end{align}
The base space is a two-dimensional Minkowski spacetime spanned by 
the coordinates $x^\mu=(t,x)$ with the metric $\eta^{\mu\nu}={\rm diag}(-1,+1)$\,. 
Then one can expand the one-form $J$ by the world-sheet coordinates as $J=J_\mu dx^\mu$. 
The rapidly dumping boundary condition is taken so that 
the group-valued field $g(x)$ approaches a constant element 
$g_{\infty}$ at spatial infinities\footnote{%
Precisely, the group-valued field depends on both time $t$ and the spatial direction $x$, namely $g=g(t,x)$\,. 
Hereafter the time dependence is suppressed like $g=g(x)$ for simplicity.}:
\begin{eqnarray}
g(x) ~\to~ g_\infty \qquad \text{with} \qquad x~ \to ~ \pm\infty \,. \label{bc}
\end{eqnarray}
That is, $J_{\mu}$ vanishes very rapidly at spatial infinities. 
This setup is not appropriate in considering some applications to string theory.  
However, it is suitable to study infinite-dimensional symmetries generated by an infinite set of 
non-local charges in a well-defined way. The Virasoro constraints are also not taken into account. 

\medskip 

Taking a variation of the action \eqref{action} leads to the equations of motion,  
\begin{eqnarray}
\partial^{\mu}J_{\mu} -2C\Tr(T^{-}J_{\mu})[J^{\mu},T^{-}]=0\,,  \label{eom2}
\end{eqnarray}
where it has already been rewritten by using the conservation law of the $U(1)_{\rm R}$ current, 
$\partial^{\mu}J^{-}_{\mu} = 0$\,. 
When $C=0$, the \sch sigma models become $SL(2,{\mathbb R})$ principal chiral models. 

\medskip 

The \sch sigma models are classically integrable \cite{KY-Sch}. 
The $SL(2,{\mathbb R})_{\rm L}$ and $U(1)_{\rm R}$ symmetries give rise to two descriptions to 
describe the classical dynamics. One is the left description based on $SL(2,{\mathbb R})_{\rm L}$\,.  
The other is the right description based on $U(1)_{\rm R}$\,. For each of them, Lax pairs and monodromy matrices 
are constructed. Then all Lax pairs lead to the identical equations of motion \eqref{eom2}. 
The classical integrable structure is similar to the hybrid one in the cases of squashed S$^3$ 
and warped AdS$_3$ \cite{KYhybrid,KMY-QAA,KMY-monodromy,KY-summary}.

\subsection{Anisotropic Lax pairs in the right description}

For later argument, let us introduce Lax pairs in the right description \cite{exotic}. 
The expressions of the Lax pairs are anisotropic to reflect the deformed target space. 
The anisotropic Lax pairs $\cA^{R_\pm}_\mu(x;\la_{R_\pm})$ are given 
by\footnote{%
We have changed the notation from \cite{exotic}, precisely 
$\cA^{R_\pm}_\mu(x;\la_{R_\pm})_{\text{[ours]}}=L^{R_\pm}_\mu(x;\la_{R_\pm}){}_{\!\text{\cite{exotic}}}$\,. 
%
As we will see 
in Section \ref{sec:Jordan}, the anisotropy of the Lax pairs is 
superficial and turns out to be isotropic in essential.}  
\begin{align}
&\cA^{R_\pm}_\mu(x;\la_{R_\pm}) = \frac{1}{1-\la_{R_\pm}^2}\bigl[
T^+(J_\mu^--\la_{R_\pm}\ep_{\mu\nu}J^{-,\nu}) \no \\
&\quad +T^2(-J_\mu^2+\la_{R_\pm} (\ep_{\mu\nu}J^{2,\nu}\pm\sqrt{C}J_\mu^-)\mp\sqrt{C}\la_{R_\pm}^2\ep_{\mu\nu}J^{-,\nu}) \no\\
&\quad +T^-\bigl(J_\mu^+ -\la_{R_\pm} (\ep_{\mu\nu}J^{+,\nu}\pm\sqrt{C}J_\mu^2+C\ep_{\mu\nu}J^{-,\nu})
+\la_{R_\pm}^2(\pm\sqrt{C}\ep_{\mu\nu}J^{2,\nu}+CJ_\mu^-) \bigr) \bigr]\,. 
\label{right lax}
\end{align}
Here $\la_{R_{\pm}}\in \mathbb{C}$ are spectral parameters
and the anti-symmetric tensor $\ep_{\mu\nu}$ is normalized as 
$\ep_{tx}=-\ep^{tx}=1$\,. 
The equations of motion \eqref{eom2} and the flatness condition of the current 
$dJ+J\wedge J=0$ 
are reproduced from the zero-curvature condition,  
\begin{align}\label{rzero} 
\bigl[\partial_t-\cA^{R_\pm}_t(x;\la_{R_\pm}), \partial_x-\cA^{R_\pm}_x(x;\la_{R_\pm})\bigr]=0\,.
\end{align}
This ensures the integrability of the systems. Indeed, the monodromy matrices 
\begin{eqnarray}
U^{R_\pm}(\lambda_{R_\pm}):= \pexp \left[\int^\infty_{-\infty}\!\!\!dx\,\cA^{R_\pm}_x(x;\lambda_{R_\pm})\right]
\end{eqnarray}
are conserved for any values of $\lambda_{R_\pm}\in \mathbb{C}$\,, 
due to the condition \eqref{rzero}\,; 
\begin{eqnarray}
\frac{d}{dt}U^{R_\pm}(\lambda_{R_\pm})=0\,.  
\end{eqnarray}
This conservation law leads to an infinite number of conserved charges.

\paragraph{General properties of Lax pairs and monodromy matrices.}

In general, Lax pairs may be regarded as gauge fields and it is transformed under gauge transformations. 
In the next section we will discuss a gauge transformation of $\cA^{R_\pm}_\mu(x;\lambda_{R_\pm})$\,.  
Before that, it is worth seeing general transformation properties of Lax pairs under gauge transformations. 

\medskip 

Let us denote an arbitrary Lax connection as $L_\mu(x;\la)$\,. 
Then the gauge transformation law is given by  
\begin{align}\label{gtrf}
\left[L_\mu(x;\la)\right]^{f(x)} :=f(x)^{-1}L_\mu(x;\la)f(x) -f(x)^{-1}\partial_\mu f(x)\,, 
\end{align}
where $f(x)$ is an arbitrary $SL(2,{\mathbb R})$-valued function of the world-sheet. 
This transformation preserves the zero-curvature condition \eqref{rzero} because 
\eqref{gtrf} can be rewritten with the covariant derivative as follows:   
\begin{align}
\label{cov}
\partial_\mu-[L_\mu(x;\la)]^{f(x)} = f(x)^{-1} \bigl(\partial_\mu-L_\mu(x;\la)\bigr)f(x)~. 
\end{align}
For any $SL(2,{\mathbb R})$-valued functions $g(x)$ and $h(x)$, 
the gauge transformation by the product $g(x)h(x)$ is equivalent to 
the product of the gauge transformations by $g(x)$ and $h(x)$: 
\begin{align} \label{composit}
[L_\mu(x;\la)]^{g(x)h(x)} = \bigl[ [L_\mu(x;\la)]^{g(x)}\bigr]^{h(x)}\,. 
\end{align}
Due to \eqref{cov}, the associated monodromy matrices are also transformed in the adjoint way,    
\begin{eqnarray}
\left[U(\la)\right]^f=f(-\infty)^{-1}U(\la)f(+\infty) \quad \text{with} \quad 
U(\la):= \pexp \left[\int^\infty_{-\infty}\!\!\!dx~L_x(x;\la)\right]\,.   \nonumber
\end{eqnarray}
The spatial boundary values $f(\pm\infty)$ 
are constant under the boundary 
condition \eqref{bc} and hence the resulting monodromy matrix is still conserved.

\subsection{Exotic symmetry}

In the present case, due to the deformation, a part of the isometry of AdS$_3$\,, 
$SL(2,\mathbb{R})_{\rm L} \times SL(2,\mathbb{R})_{\rm R} $ is broken to 
$SL(2,\mathbb{R})_{\rm L} \times U(1)_{\rm R}$\,. 
Here we shall focus upon the breaking of $SL(2,\mathbb{R})_{\rm R}$ to $U(1)_{\rm R}$\,. 
The Noether current associated with $U(1)_{\rm R}$ is given by 
\begin{align} \label{Jminus}
j^{R,-}_\mu=-J^-_\mu\,. 
\end{align}
As the isometry of the target space, the $2,+$ components of $\alg{sl}(2)_{\rm R}$ are broken. 
However, as the symmetry of the sigma models, there still exist the conserved charges 
corresponding to these components. Those are realized in a non-local way and 
the conserved currents are given by \cite{KY-Sch, exotic}  
\begin{eqnarray}
\label{j2p}
&&j^{R,2}_\mu=-{\rm e}^{\sqrt{C}\chi^-}(J^2_\mu+\sqrt{C}\ep_{\mu\nu}J^{\nu,-})\,, \no\\ 
&&j^{R,+}_\mu=-{\rm e}^{\sqrt{C}\chi^-}(J^+_\mu+\sqrt{C}\ep_{\mu\nu}J^{\nu,2}+CJ^-_\mu)\,. 
\end{eqnarray}
Here $\chi^-(x)$ is the non-local field defined as 
\begin{align}\label{chi}
\chi^-(x):=-\frac{1}{2}\int^\infty_{-\infty}\!\!\!dy~\ep(x-y)j^{R, -}_t(y)\,, 
\end{align}
and $\ep(x)$ is the signature function defined as $\ep(x) := \theta(x) - \theta(-x)$ 
with the step function $\theta(x)$\,. 
The associated conserved charges are given by 
\begin{align}
Q^{R,a}=\int^\infty_{-\infty}\!\!\!dx\,j^{R,a}_t(x)
\qquad (a=\pm,2)\,.
\end{align}
Note that the non-local descriptions are intrinsic to the sigma model realizations rather than the isometry. 

\medskip 

In addition to \eqref{j2p}, 
there is another set of conserved 
currents $\wt{j}^{R,a}_\mu$ represented by 
\begin{eqnarray}
&&\wt{j}^{R,-}_\mu=j^{R,-}_\mu=-J^-_\mu\,, \nonumber \\
&&\wt{j}^{R,2}_\mu=-{\rm e}^{-\sqrt{C}\chi^-}(J^2_\mu-\sqrt{C}\ep_{\mu\nu}J^{\nu,-})\,, \,~ \nonumber \\
&&\wt{j}^{R,+}_\mu=-{\rm e}^{-\sqrt{C}\chi^-}(J^+_\mu-\sqrt{C}\ep_{\mu\nu}J^{\nu,2}+CJ^-_\mu)\,. 
\end{eqnarray}
The currents $\wt{j}^{R,a}_\mu$ are related to $j^{R,a}_\mu$ by flipping the signature of $\sqrt{C}$\,,  
where $j^{R,-}_{\mu}$ does not contain $\sqrt{C}$ and hence 
$\wt{j}^{R,-}_\mu$ coincides with $j^{R,-}_\mu$\,. 
The associated charges 
\begin{eqnarray} 
\wt{Q}^{R,a}=\int^\infty_{-\infty}\!\!\!dx~\wt{j}^{R,a}_t(x) \qquad (a=\pm,2)
\end{eqnarray}
are also conserved. 

\medskip 

The Poisson algebra of $Q^{R,a}$ is computed in \cite{KY-Sch} 
by imposing the canonical commutation relations among the dynamical variables 
$X(x)\in\{v(x),\rho(x), u(x)\}$ and the conjugate momenta, 
\begin{align}
\bigl\{X(x),\Pi_X(y) \bigr\}_{\rm P}=\de(x-y) \qquad \text{with}\qquad 
\Pi_X(y) =\frac{\partial L}{\partial \dot X}(y)\,.  
\end{align}
The dot means time-derivative $\dot X=\partial_t X$. 
The resulting algebra turns out to be the classical analogue of 
$q$-deformed two-dimensional Poincar\'e algebra \cite{q-Poincare,Ohn};  
\begin{align}
&\{Q^{R,2},Q^{R,+}\}_{\rm P}=Q^{R,+}\cosh(\xi Q^{R,-})\,, \quad 
\{Q^{R,2},Q^{R,-}\}_{\rm P}=-\frac{\sinh(\xi Q^{R,-})}{\xi}\,, \no\\
&\{Q^{R,-},Q^{R,+}\}_{\rm P}=Q^{R,2}\,.
\label{qpo}
\end{align}
Note that the deformation parameter has been redefined from $C$ to $\xi$ as 
\begin{eqnarray}
\xi:=\frac{\sqrt{C}}{2}\,.
\end{eqnarray} 
This algebra is also referred as a non-standard $q$-deformation of $\cU(\alg{sl}(2))$ 
with $q={\rm e}^\xi$ in comparison to the standard quantum deformation 
$\cU_q(\alg{sl}(2))$ \cite{Drinfeld1, Drinfeld2, Jimbo} (see also sec.\,6.4\,F in \cite{CP}).  
Since the relations in \eqref{qpo} are invariant under the sign flip $\xi\to -\xi$\,,  
another set of the charges $\wt{Q}^{R,a}$ also generates a copy of the $q$-Poincar\'e algebra. 

\medskip 

Notably, taking the mixed commutators $\{Q^{R,a},\wt{Q}^{R,b}\}_{\rm P}$ into account, 
the whole algebra generates an infinite-dimensional symmetry. 
That is, the charges with tilde can be regarded as {\it affine} 
generators\footnote{In the case of squashed spheres, the same technique leads to a quantum affine algebra 
$\cU_q(\widehat{\alg{sl}}(2))$ \cite{KMY-QAA}.}. We refer to this symmetry 
as {\it exotic symmetry} \cite{exotic}. 
It is remarkable that the exotic symmetry has only semi-infinite levels as well as the Yangians.  
Even though the commutation relations among the charges are not completely determined yet,  
the following relation is enough for our later argument, 
\begin{align}\label{QQtl}
\{\wt{Q}^{R,2}, Q^{R,2}\}_{\rm P}=\cosh(\xi Q^{R,-}) (\wt{Q}^{R,2} -Q^{R,2})\,. 
\end{align}

\section{Gauge transformations and Yangian \texorpdfstring{$\cY(\alg{sl}(2))$}{Y(sl2)} 
\label{sec:Jordan}}

In this section we shall rewrite the exotic symmetry into the undeformed Yangian  $\cY(\alg{sl}(2))$ 
by performing a gauge transformation.  

\medskip 

Let us first summarize some properties of the exotic symmetry. The list of them gives rise to an important hint 
to find out the gauge transformation. The properties are the following: 
\begin{enumerate}
\item  The charge algebra of the exotic symmetry exhibits the semi-infinite levels such as 
Yangians (For the concrete shape of root lattice, see \cite{exotic}). 
\vspace{-3mm}
\item  The spectral parameter relations in the left-right duality 
is the same as the ones in $SL(2)$ principal chiral models.  
\vspace{-3mm}
\item  The classical $r/s$-matrices associated with the exotic symmetry suggest that 
the right description also belongs to the rational class\footnote{%
Due to the presence of non-ultra local terms, 
this classification is not definite though.}.
\end{enumerate}
These properties strongly suggest that the isotropic Lax pairs should exist in the right description. 

\medskip 

If this observation is true, then there should be the gauge transformations between the anisotropic Lax pairs $\cA^{R_\pm}_\mu(x;\la_{R_\pm})$ 
in \eqref{right lax} and the isotropic ones. 
Then the isotropic Lax pairs are expected to 
take the following form 
\begin{align}
\cL^{R_\pm}_\mu(x;\la_{R_\pm})=\frac{\cJ_\mu^{R_\pm}-\la_{R_\pm}\ep_{\mu\nu}\cJ^{R_\pm, \nu}}{1-\la_{R_\pm}^2}\,, 
\label{nonlocal lax}
\end{align}
together with a conserved current $\cJ^{R_\pm}_\mu(x)$ which satisfies the flatness condition. 

\medskip 

We will construct concretely the isotropic Lax pairs with flat conserved currents from the anisotropic ones  
by performing non-local gauge transformations below. 
Then we present the current algebras for the flat conserved currents. 
The BIZZ procedure \cite{BIZZ} is applicable to generate an infinite number of conserved charges. 
The resulting charge algebras are shown to be the undeformed Yangian $\cY(\alg{sl}(2))$\,. 
The classical $r$/$s$-matrices for the isotropic Lax pairs are also computed by following the work \cite{Maillet}. 
Finally we discuss the relation between the exotic symmetry and the undeformed Yangian.

\subsection{Non-local gauge transformations \label{sec:nonlocalGT} }

The first we have to do is to find out a gauge transformation from the anisotropic Lax pair 
to the desired isotropic ones. 

\medskip 

To get a hint, let us compare the asymptotic behavior around $\la_{R_\pm}= \infty$\,. 
The isotropic Lax pairs $\cL^{R_\pm}_\mu(x;\la_{R_\pm})$ are expected to vanish at $\la_{R_{\pm}}=\infty$ from \eqref{nonlocal lax}\,. 
On the other hand, the asymptotic behavior of the anisotropic Lax pairs $\cA^{R_\pm}_\mu(x;\la_{R_\pm})$ is fixed 
from the concrete expressions. It gives rise to the ${\cal O}(1)$ contributions at $\la_{R_{\pm}}=\infty$ 
and the asymptotic forms are given by 
\begin{align}
\cA^{R_\pm}_\mu(x;\infty)=
\pm\sqrt{C}\bigl[T^2\ep_{\mu\nu}J^{-,\nu}-T^-(\ep_{\mu\nu}J^{2,\nu}\pm\sqrt{C}J^-_\mu)\bigr]\,.  
\label{remaining}
\end{align}
At this point, one should notice that the Lax pairs that vanishes at $\la_{R_\pm}= \infty$ can be constructed from the original $\cA^{R_\pm}_\mu(x;\la_{R_\pm})$ 
by subtracting the remaining contribution (\ref{remaining})\,. Then this subtraction is indeed realized as a gauge transformation, 
as we will see below. 

\medskip 

The next is to consider the gauge transformation to realize the subtraction. 
To define the gauge transformation it is necessary to find out an appropriate 
gauge functions. Indeed, the gauge functions are given by the following {\it non-local} fields 
\begin{align}\label{twist}
\cF^\pm(x):= \pexp \Bigl[\int^x_{-\infty}\!\!\!dy~\cA^{R_\pm}_x(y;\infty)\Bigr] K^\pm\,,  
\end{align}
where $K^\pm$ are arbitrary constant $SL(2,{\mathbb R})$ elements. 
Those are interpreted as formal solutions of the differential equation 
\begin{align}\label{difeq}
[\partial_\mu-\cA^{R_\pm}_\mu(x;\infty)] \cF^\pm(x)=0\,, 
\end{align}
and $K^{\pm}$ corresponds to the overall constant factor of the solutions. 

\medskip 

The gauge transformations of $\cA^{R_\pm}_\mu(x;\la_{R_\pm})$ by $\cF^\pm(x)$ are performed as follows: 
\begin{align}
[\cA^{R_\pm}_\mu(x;\la_{R_\pm})]^{\cF^\pm(x)} 
&=\cF^\pm(x)^{-1}\cA^{R_\pm}_\mu(x;\la_{R_\pm})\cF^\pm(x) - \cF^\pm(x)^{-1}\partial_\mu \cF^\pm(x) \no \\
&=\cF^\pm(x)^{-1}\bigl(\cA^{R_\pm}_\mu(x;\la_{R_\pm}) - \cA^{R_\pm}_\mu(x;\infty)\bigr)\cF^\pm(x)~. 
\end{align}
In the second equality the differential equation \eqref{difeq} has been used. 
Now it is obvious that $[\cA^{R_\pm}_\mu(x;\la_{R_\pm})]^{\cF^\pm(x)} $
vanish at $\la_{R_\pm}\to \infty$\,. Furthermore, as a byproduct, it turns out that 
$[\cA^{R_\pm}_\mu(x;\la_{R_\pm})]^{\cF^\pm(x)} $ take the desired isotropic forms, 
\begin{align}
\cL^{R_\pm}_\mu(x;\la_{R_\pm}):=[\cA^{R_\pm}_\mu(x;\la_{R_\pm})]^{\cF^\pm(x)} 
=\frac{\cJ^{R_\pm}_\mu-\la_{R_\pm}\ep_{\mu\nu}\cJ^{R_\pm, \nu}}{1-\la_{R_\pm}^2}\,.   
\label{rflatlax}
\end{align}
Here the currents $\cJ_\mu^{R_\pm}$ are defined as   
\begin{align}\label{rflatcur}
\cJ^{R_\pm}_\mu(x):= -(g\cF^\pm(x))^{-1}\partial_\mu (g\cF^\pm(x))\,, 
\end{align}
which satisfy the flatness condition by definition. 
The conservation law can also be shown by direct computation. 

\medskip 

Here we should be careful of the ambiguity of the constant factors $K^\pm$ in \eqref{twist}.  
This ambiguity may be regarded as a kind of residual gauge freedom. 
The reason is as follows. 
If we take the gauge-fixing condition as 
\begin{eqnarray}
\cL^{R_\pm}_\mu(x;\lambda_{R_\pm})\to 0
\qquad {\rm as} \qquad \lambda_{R_\pm}\to \infty\,, 
\end{eqnarray}
then the gauge potential $\cF^\pm$ are not completely fixed 
and then there is still ambiguity of the overall constant factor. 
This remaining freedom of the gauge transformation is identified with the choice of $K^\pm$\,. 

\medskip 

In fact, any choice of $K^{\pm}$ works well as the isotropic Lax pairs by the construction. 
However, the expressions of the associated currents are sensitive to the choice. 
To see this, let us consider the right multiplication for $K^\pm$ 
by a constant matrix element $H$.   
Then the currents $\cJ^{R_\pm}_\mu$ are transformed in the adjoint way, namely 
\begin{eqnarray}
\cJ^{R_\pm}_\mu \to H^{-1}\cJ^{R_\pm}_\mu H \qquad \text{under} \qquad
K^\pm \to K^\pm H\,. 
\end{eqnarray}
This transformation triggers the mixing among the $\alg{sl}(2)$ components of $\cJ^{R_\pm}_\mu$\,. 
In particular, the resulting form of the current algebra (as well as the charge algebra) depends on the mixing.  
This fact implies the possibility that the exotic symmetry can be rewritten into some known algebra 
by a suitable gauge fixing of $H$\,. This is the key observation in our argument. 

\medskip 

Although we are going to discuss the current algebra hereafter, before that, 
we have to fix the remaining gauge freedom $K^{\pm}$\,. In doing that, 
we would like to take the gauge that leads to the simplest form of the current algebra. 
How can we do that? What are hints to do that? The important clue is provided by 
the mathematical background on quantum Jordanian twists. The observation to find out 
how to fix the gauge is explained in the next section \ref{sec:Jordanian}\,. 
Here we shall simply give the answer; 
\begin{align}\label{Kfix}
K^\pm =\exp (\pm2\xi T^2 Q^{R,-})  
=\biggl( \; \begin{matrix} {\rm e}^{\pm\xi Q^{R, -}}  & 0 \\ 0 & {\rm e}^{\mp\xi Q^{R, -}} \; \end{matrix} \biggr)\,. 
\end{align}
This gauge fixing leads to the undeformed Yangian, as we will see later.
We also discuss an example of the different gauge-fixing in Appendix \ref{app:gfix}. 

\medskip 

The next is to show the current algebra under the choice (\ref{Kfix})\,. 
First of all, let us derive the expressions of the currents.  
The non-local fields $\cF^\pm$ are given by 
\begin{align} \label{Fpm}
\cF^+(x)&= \exp\bigl[-2\xi T^- {\rm e}^{-2\xi \chi^-(x)}(\chi^2(x)-\tfrac{1}{2}Q^{R,2})\bigr] 
\exp\bigl[+2\xi T^2(\chi^-(x)+\tfrac{1}{2}Q^{R,-})\bigr]\,, \no\\ 
\cF^-(x)&= \exp\bigl[+2\xi T^- {\rm e}^{+2\xi \chi^-(x)}(\wt{\chi}^2(x)-\tfrac{1}{2}\wt{Q}^{R,2})\bigr] 
\exp\bigl[-2\xi T^2(\chi^-(x)+\tfrac{1}{2}Q^{R,-})\bigr]\,. 
\end{align}
Then $\cJ^{R_\pm}_\mu$ are given by 
\begin{align}
\cJ_\mu^{R_+,+}(x)&={\rm e}^{+\xi Q^{R,-}} j^{R,+}_\mu(x) 
+ \xi \ep_{\mu\nu}\partial^\nu \bigl[{\rm e}^{-2\xi(\chi^-(x) 
-\tfrac{1}{2}Q^{R,-})}(\chi^2(x)-\tfrac{1}{2}Q^{R,2})^2 \bigr]\,, \no\\
\cJ_\mu^{R_-,+}(x)&={\rm e}^{-\xi Q^{R,-}} \wt{j}^{R,+}_\mu(x) 
- \xi \ep_{\mu\nu}\partial^\nu \bigl[{\rm e}^{+2\xi(\chi^-(x) 
-\tfrac{1}{2}Q^{R,-})}(\wt{\chi}^2(x)-\tfrac{1}{2}\wt{Q}^{R,2})^2 \bigr]\,, \no\\
\cJ_\mu^{R_+,2}(x)&=-\ep_{\mu\nu}\partial^\nu 
\bigl[{\rm e}^{-2\xi\chi^-(x)}(\chi^2(x)-\tfrac{1}{2}Q^{R,2}) \bigr]\,, \no\\
\cJ_\mu^{R_-,2}(x)&=-\ep_{\mu\nu}\partial^\nu 
\bigl[{\rm e}^{+2\xi\chi^-(x)}(\wt{\chi}^2(x)-\tfrac{1}{2}\wt{Q}^{R,2}) \bigr]\,, 
\no\\ 
\cJ_\mu^{R_+,-}(x)&=+\tfrac{1}{2\xi} \ep_{\mu\nu}\partial^\nu 
\bigl[{\rm e}^{-2\xi (\chi^-(x)+\tfrac{1}{2}Q^{R,-})}\bigr]\,,  \no \\
\cJ_\mu^{R_-,-}(x)&=-\tfrac{1}{2\xi} \ep_{\mu\nu}\partial^\nu 
\bigl[{\rm e}^{+2\xi (\chi^-(x)+\tfrac{1}{2}Q^{R,-})}\bigr]\,.    
\label{rflatexp} 
\end{align}
Here the non-local field $\chi^-(x)$ is given in \eqref{chi}. 
New non-local fields $\chi^2(x)$ and $\wt{\chi}^2(x)$ 
are defined as, respectively,  
\begin{eqnarray}
\chi^2(x) :=-\frac{1}{2}\int^\infty_{-\infty}\!\!\!dy\,\epsilon(x-y)j^{R,2}_t(y)\,, \quad
\wt{\chi}^2(x) :=-\frac{1}{2}\int^\infty_{-\infty}\!\!\!dy\,\epsilon(x-y)\wt{j}^{R,2}_t(y)\,.
\end{eqnarray}
For the derivations of (\ref{Fpm}) and (\ref{rflatexp}), see Appendix \ref{app:twistF}. 

\medskip 

The currents $\cJ^{R_\pm}_\mu$ are quite complicated, but the current algebra takes a simple form, 
\begin{align}
\bigl\{\cJ^{R_\pm, a}_t(x), \cJ^{R_\pm, b}_t(y)\bigr\}_{\rm P}
&= \ve^{ab}{}_c\cJ^{R_\pm, c}_t(x)\delta(x-y)\,,  \no \\
\bigl\{ \cJ^{R_\pm, a}_t(x), \cJ^{R_\pm, b}_x(y)\bigr\}_{\rm P}
&=\ve^{ab}{}_c \cJ^{R_\pm, c}_x(x)\delta(x-y) +\ga^{ab}\partial_x\delta(x-y)\,, \no \\
\bigl\{ \cJ^{R_\pm, a}_x(x), \cJ^{R_\pm, b}_x(y)\bigr\}_{\rm P}
&= 0\,. \label{ca} 
\end{align}
The detail of the computations is explained in Appendix \ref{ca:app}. Notably, this current algebra 
is exactly the same as the one in the usual $SL(2,{\mathbb R})$ principal chiral models. 

\subsection{The undeformed Yangian algebra}

The next is to consider the charge algebra generated by the current $\cJ^{R_\pm}_\mu$\,. 

\medskip 

First of all, let us derive a set of the conserved charges. 
The current $\cJ^{R_\pm}_\mu$ is non-local but satisfies the flatness condition by the definition \eqref{rflatcur}. 
Thus one can generate an infinite number of conserved charges by applying  the BIZZ construction \cite{BIZZ}. 
The first few charges are inductively obtained as    
\begin{align}
\cY^{R_\pm}_{(0)}&=\int^\infty_{-\infty}\!\!\!dx~\cJ^{R_\pm}_t(x)\,, \no \\
\cY^{R_\pm}_{(1)}&=\frac{1}{4}\int^\infty_{-\infty}\!\!\!dx\int^\infty_{-\infty}\!\!\!dy~
\ep(x-y)\left[\cJ^{R_\pm}_t(x),\cJ^{R_\pm}_t(y)\right]-\int^\infty_{-\infty}\!\!\!dx~\cJ^{R_\pm}_x(x)\,, \no \\
\cY^{R_\pm}_{(2)}&=\frac{1}{12}\int^\infty_{-\infty}\!\!\!dx\int^\infty_{-\infty}\!\!\!dy\int^\infty_{-\infty}\!\!\!dz~
\ep(x-y)\ep(x-z)\left[\bigl[\cJ^{R_\pm}_t(x),\cJ^{R_\pm}_t(y)\bigr],\cJ^{R_\pm}_t(z)\right] \nonumber \\
&\quad -\frac{1}{2}\int^\infty_{-\infty}\!\!\!dx\int^\infty_{-\infty}\!\!\!dy~
\ep(x-y)\left[\cJ^{R_\pm}_t(x),\cJ^{R_\pm}_x(y)\right]+\int^\infty_{-\infty}\!\!\!dx~\cJ^{R_\pm}_t(x)\,. 
\label{right-bizz}
\end{align} 
In principle, one can derive the expression with an arbitrary order. 

\medskip 

Then the current algebra \eqref{ca} leads to 
the Poisson brackets of $\cY^{R_\pm,a}_{(0)}$ and $\cY^{R_\pm,a}_{(1)}$\,,  
\begin{align}
\bigl\{ \cY^{R_\pm,a}_{(0)}, \cY^{R_\pm,b}_{(0)}\bigr\}_{\rm P}&=\ve^{ab}_{~~c} \cY^{R_\pm,c}_{(0)}\,, \no \\
\bigl\{ \cY^{R_\pm,a}_{(1)}, \cY^{R_\pm,b}_{(0)}\bigr\}_{\rm P}&=\ve^{ab}_{~~c} \cY^{R_\pm,c}_{(1)}\,, \no \\
\bigl\{ \cY^{R_\pm,a}_{(1)}, \cY^{R_\pm,b}_{(1)}\bigr\}_{\rm P}&=\ve^{ab}_{~~c} \bigl[ \cY^{R_\pm,c}_{(2)}
+\tfrac{1}{12} ( \cY^{R_\pm}_{(0)} )^{2} \cY^{R_\pm,c}_{(0)} -\cY^{R_\pm,c}_{(0)} \bigr]\,, 
\label{level2}
\end{align} 
up to a subtlety of non-ultra local terms\footnote{In computing the algebra, 
we have followed a prescription utilized in \cite{MacKay}.}. 
These are nothing but the defining relations of the Yangian algebra \cite{Drinfeld1, Drinfeld2}. 
Indeed, after some algebra, the Serre relations are also derived,  
\begin{align}
\bigl\{\bigl\{\cY^{+}_{(1)},\cY^{-}_{(1)}\bigr\}_{\rm P},\cY^{2}_{(1)}\bigr\}_{\rm P}
=\tfrac{1}{4}(\cY^{+}_{(1)}\cY^{-}_{(0)}-\cY^{+}_{(0)}\cY^{-}_{(1)})\cY^{2}_{(0)}\,.  
\end{align} 
Thus the set of the conserved charges $\cY^{R_\pm,a}_{(n)}$ ($n\geq 0$) generate the Yangian algebra 
$\cY(\alg{sl}(2))$ in the sense of Drinfeld's first realization \cite{Drinfeld1, Drinfeld2}. 

\medskip 

The charges can also be produced by expanding $U^{R_\pm}(\la_{R_\pm})$ around $\la_{R_\pm}=\infty$ like 
\begin{align}
\bigl[U^{R_\pm}(\la_{R_\pm})\bigr]^{\cF^\pm}=\cF^\pm(+\infty)^{-1} U^{R_\pm}(\la_{R_\pm}) \cF^\pm(-\infty) 
=\exp \Bigl[\sum_{n=0}^\infty\cY^{R_\pm}_{(n)} \la_{R_\pm}^{-n-1} \Bigr]~.
\label{monexp}
\end{align}
Note that the gauge transformation plays an important role in finding out the Yangian charges properly. 
If we naively expand the original $U^{R_\pm}(\la_{R_\pm})$ around $\la_{R_\pm}=\infty$\,, 
then some messy expressions, 
which cannot be identified with the Yangian charges at a glance,  would be obtained.

\medskip 

So far, the two sets of Yangians charges  $\cY^{R_+}_{(n)}$ and $\cY^{R_-}_{(n)}$ have been constructed 
from $U^{R_+}(\lambda_{R_+})$ and $U^{R_-}(\lambda_{R_-})$ respectively. 
Note that $\cY^{R_-}_{(n)}$ does not coincide with $\cY^{R_+}_{(n)}$. 
This is contrast to the two sets of Yangians in the left description, where the currents contain 
the deformation parameter $\xi=\sqrt{C}/2$ while the dependence does not appear at the charge 
level\footnote{For the level-zero charges, the terms proportional to $\xi$ are integrated out (see \eqref{improved}). 
For the level-one charges, the integrated forms contain the $\xi$-dependent terms 
which are linear to the level-zero charges. Such terms can be removed by using the 
automorphism of Yangian. After all, there is no $\xi$-dependence.}.
Summarizing, the two Yangians are degenerate in the left description 
while it is not the case in the right description. 

\medskip 

Before closing this section, let us examine the $r$/$s$-matrices for ${\mathcal L}^{R_\pm}_\mu(x;\la_{R_\pm})$ 
by following the work \cite{Maillet}. With the current algebra \eqref{ca}, the following bracket is evaluated as  
\begin{align}
&\left\{ \cL^{R_\pm}_x(x;\la_{R_\pm})\stackrel{\otimes}{,} \cL^{R_\pm}_x(y;\mu_{R_\pm})\right\}_{\rm P} \no \\
=&\left[r^{R_\pm}(\la_{R_\pm},\mu_{R_\pm})-s^{R_\pm}(\la_{R_\pm},\mu_{R_\pm}), \;
\cL^{R_\pm}_x(x;\la_{R_\pm})\otimes 1\right]\delta(x-y) \no \\
&+ \left[r^{R_\pm}(\la_{R_\pm},\mu_{R_\pm})+s^{R_\pm}(\la_{R_\pm},\mu_{R_\pm}), \;
1\otimes \cL^{R_\pm}_x(y;\mu_{R_\pm})\right]\delta(x-y) \no \\
&-2s^{R_\pm}(\la_{R_\pm},\mu_{R_\pm})\partial_x\delta(x-y)\,. 
\end{align}
Here the $r$-matrices $r^{R_\pm}(\la_{R_\pm},\mu_{R_\pm})$ and 
$s$-matrices $s^{R_\pm}(\la_{R_\pm}, \mu_{R_\pm})$ are given by, respectively,  
\begin{align}
r^{R_\pm}(\la_{R_\pm},\mu_{R_\pm})
&=\frac{1}{2 (\la_{R_\pm}-\mu_{R_\pm} )}
\Bigl(\frac{\la_{R_\pm}^2}{1-\la_{R_\pm}^2}+\frac{\mu_{R_\pm}^2}{1-\mu_{R_\pm}^2}\Bigr) 
\ga_{ab} T^a\otimes T^b  \,, \no \\
s^{R_\pm}(\la_{R_\pm},\mu_{R_\pm})
&=\frac{\la_{R_\pm}+\mu_{R_\pm}}{2 (1-\la_{R_\pm}^2 ) (1-\mu_{R_\pm}^2 )} \ga_{ab} T^a\otimes T^b\,.   
\label{r/s-matrices}
\end{align} 
These are exactly the same as those in the $SL(2,\mathbb{R})$ principal chiral models.

\subsection{The relation to the exotic symmetry \label{sec:YtoQ} }

It is of importance to elucidate the relation between the original exotic symmetry 
and the Yangian $\cY(\alg{sl}(2))$ that is obtained after the gauge transformation.  
The charges at the level-zero are expressed in terms of the $q$-Poincar\'e generators $Q^{R,a}$ as follows: 
\begin{align}
\cY^{R_+,+}_{(0)}&= {\rm e}^{\xi Q^{R,-}}Q^{R,+}+\xi  \bigl({\rm e}^{\xi Q^{R,-}} Q^{R,2}\bigr)^2\,, \no \\
\cY^{R_+,2}_{(0)}&= {\rm e}^{\xi Q^{R,-}} Q^{R,2}\,, \no \\ 
\cY^{R_+,-}_{(0)}&=\frac{1-{\rm e}^{-2\xi Q^{R,-}}}{2\xi }\,. 
\label{level0} 
\end{align}
These relations are obtained from the relations between 
$\cJ^{R_+}_\mu$ and $j^R_\mu$ in 
\eqref{rflatexp}\,. 
Note that $\cY^{R_+,a}_{(0)}$'s form the undeformed $\cU(\alg{sl}(2))$\,. 
In the mathematical language, the relations in \eqref{level0} are regarded as a 
homomorphism from $\cU(\alg{sl}(2))$ to the $q$-deformed Poincar\'e algebra \eqref{qpo} 
at the classical level\footnote{%
We expect that an appropriate completion makes this map the isomorphism
between $\cU(\alg{sl}(2))$ and the $q$-Poincar\'e algebra as the Poisson algebra.}. 
To express the level-one charge, we also need 
not only $q$-deformed Poincar\'e charges but also the non-trivial exotic charges 
$\wt{Q}^{R,2}$ and $\wt{Q}^{R,+}$ like 
\begin{align}
\cY^{R_+,+}_{(1)}&=\tfrac{1}{4} {\rm e}^{\xi Q^{R,-}} \bigl[
\tfrac{1}{\xi}\{\{\wt{Q}^{R,2}, Q^{R,+}\}_{\rm P}, Q^{R,+}\}_{\rm P} +  \{\wt{Q}^{R,2}, Q^{R,+}\}_{\rm P} Q^{R,2} \no \\
&\quad \qquad \qquad -(\sinh(\xi Q^{R,-})\tilde Q^{R,2}+\cosh(\xi Q^{R,-}) Q^{R,2})Q^{R,2} \bigr]\,, \no \\
\cY^{R_+,2}_{(1)}&=\tfrac{1}{4\xi }\bigl(\{\wt{Q}^{R,2}, Q^{R,+}\}_{\rm P}-\cosh(\xi Q^{R,-})Q^{R,+}\bigr)
+\tfrac{1}{4}Q^{R,2}(\wt{Q}^{R,2} -Q^{R,2})\,,  \no \\
\cY^{R_+,-}_{(1)}&=\tfrac{1}{4\xi}{\rm e}^{-\xi Q^{R,-}}(\wt{Q}^{R,2} -Q^{R,2})\,.   
\label{level1}
\end{align}
The defining relations of the Yangian $\cY(\alg{sl}(2))$ can be shown 
with the algebra of the exotic symmetry in \eqref{QQtl}.  
This is an alternative derivation of the Yangian $\cY(\alg{sl}(2))$ without the current algebra 
of the flat conserved current. 
The relations in \eqref{level0} and \eqref{level1} are also regarded as a homomorphism 
from the Yangian $\cY(\alg{sl}(2))$ to the exotic symmetry \cite{exotic} 
as the classical Poison algebra. 
It is an interesting issue to seek for the quantum-level expressions of 
the maps \eqref{level0} and \eqref{level1} with the Hopf algebraic structure. 

\medskip 

Note that the $\xi\to0 $ limit of \eqref{level0} and \eqref{level1} is non-singular 
and the charges reduce to the Yangian algebra realized in the usual way  
in $SL(2,{\mathbb R})$ principal chiral models. 

\medskip 

Now we have the two sets of Yangians $\cY^\pm(\alg{sl}(2))$\,. 
For instance, the level-zero charges in $\cY^-(\alg{sl}(2))$ are the same as those in \eqref{level0}\,,  
up to the replacement $(Q^R,\xi) \to (\wt{Q}^R,-\xi)$\,.  
Either of the two Yangians is a subalgebra of the exotic symmetry through 
the maps \eqref{level0} and \eqref{level1}.  
An interesting question is whether the direct sum of two Yangians  
$\cY^+(\alg{sl}(2))\oplus \cY^-(\alg{sl}(2))$ is isomorphic to the exotic symmetry or not. 
For the moment, it would be difficult to answer it because the defining relations of  
the exotic symmetry has not completely been fixed yet. 
Presumably, one can get a clue by comparing them to the relations of the deformed Yangian 
$\cY_\xi(\alg{sl}(2))$ \cite{KST} and its classical limit.

\section{The gauge fixing from Jordanian twists 
\label{sec:Jordanian}} 

The gauge fixing of the constant part $K^\pm$ in \eqref{Kfix} 
was given in an a priori way in the previous section\footnote{%
An example of the different 
gauge-fixing is discussed in Appendix \ref{app:gfix}.}. 
The aim here is to explain how $K^{\pm}$ in \eqref{Kfix} 
can be figured out from the mathematical background 
on quantum Jordanian twists. 
We first introduce the twisting procedure for the quantum case \cite{Drinfeld1, Drinfeld2, R, KST}. 
Then we compare it to the gauge transformation as the classical counter-part 
and determine the explicit forms of $K^{\pm}$\,.

\subsection{Quantum Jordanian twists \label{sec:qJordan} }

Let us recall the definition of the quasitriangular Hopf algebra introduced by Drinfeld \cite{Drinfeld1, Drinfeld2}. 
A triplet $(\cA,\copro, \cR)$, consisting of a Hopf algebra $\cA$ with the coproduct 
$\copro: \cA\to \cA\otimes \cA $ and an invertible element $\cR\in\cA\otimes \cA$\,,  
is called the quasitriangular Hopf algebra if $\cR$ satisfies the following properties: 
\begin{align}
\widetilde{\Delta}(a)\cR & =\cR\Delta(a)\qquad \text{for any } \quad a\in \cA\,, \no \\
(\Delta\otimes1)\cR &=\cR_{13}\cR_{23}\,,   
\qquad (1\otimes \Delta)\cR =\cR_{13}\cR_{12}\,, 
\label{qtr}
\end{align}
where $\coproop:=P_{12}\circ \copro$ with the permutation operator 
$P_{12}:\cA\otimes \cA\mapsto \cA\otimes \cA, P_{12}(x\otimes y)=y\otimes x$
for any $x,y\in \cA$\,. 
We denote $\cR_{12}=\cR\otimes1, \cR_{13}=\sum_i a_i\otimes1\otimes b_i, \cR_{23}=1\otimes\cR$
with $\cR=\sum_i a_i\otimes b_i$\,. 
As an important proposition, the universal R-matrix $\cR$ automatically 
satisfies the Yang-Baxter equation 
\begin{eqnarray}
\cR_{12}\cR_{13}\cR_{23}=\cR_{23}\cR_{13}\cR_{12}\,, 
\label{YBE}
\end{eqnarray} 
due to the properties \eqref{qtr}. 
Thus quasitriangular Hopf algebra is the important structure behind quantum integrable models.  

\medskip 

The quasitriangular Hopf algebra allows integrable deformations called {\it twists} \cite{Drinfeld1,Drinfeld2, R}. 
Let us suppose the existence of an invertible element $\cF\in \cA\otimes \cA$ satisfying 
the cocycle condition:
\begin{align}
\label{cocy}
\cF_{12}(\copro\otimes1)\cF=\cF_{23}(1\otimes\copro)\cF\,,    
\end{align}
where $\cF_{12}=\cF\otimes1$ and $\cF_{23}=1\otimes \cF$. 
With this twist operator, the twisted coproduct and the universal R-matrix 
can be introduced as 
\begin{align} 
\label{twistedR} 
\copro^{(\cF)}(a):=\cF\copro(a) \cF^{-1}\,, \qquad \cR^{(\cF)}:=\cF_{21} \cR \cF^{-1} \quad 
\mbox{with}~~ \cF_{21}=P_{12}\circ \cF\,.
\end{align}
Then the triplet $(\cA, \copro^{(\cF)}, \cR^{(\cF)})$ is also a quasitriangular Hopf algebra.  
As a result, the twisted R-matrix $\cR^{(\cF)}$ also satisfies the Yang-Baxter equation. 
The twist is referred as {\it the Reshetikhin twist} or {\it the Jordanian twist} 
if it is abelian $[\cF_{ij}, \cF_{kl}]=0$ or non-abelian $[\cF_{ij}, \cF_{kl}]\neq0$\,, respectively. 

\medskip 

We are interested here in a deformation of Yangian $\cA=\cY(\alg{sl}(2))$
by the {\it Jordanian} twist operator \cite{GGS}
\begin{align}
\cF&=\exp\bigl(\tfrac{1}{2}\gen{h}\otimes \ln \ga \bigr) \no \\
&=1+\sum_{n=1}^{\infty} \frac{\xi^n}{n!} \prod_{l=0}^{n-1} (\gen{h}+2l) \otimes \gen{f}^n 
=1+\xi \gen{h}\otimes \gen{f} + \tfrac{\xi^2}{2}\gen{h}(\gen{h}+2)\otimes \gen{f}^2 +\cdots\,,  
\label{F}
\end{align}
where $\ga:=1-2\xi \gen{f}$ and $\xi\in \mathbb{C}$ is a deformation parameter. 
The generators $\gen{e}, \gen{h}, \gen{f}$ are the canonical basis of $\alg{sl}(2)$\,. 
There is a long history in studies of the Jordanian deformations in quantum integrable models. 
For the related works, for example, see \cite{GGS, KST, KLM, BH, KS, Zak}. 
Note that the twist becomes trivial $\cF\to1$ in $\xi\to0$ limit.  
The twist operator $\cF$ in \eqref{F} satisfies the cocycle condition \eqref{cocy}\footnote{%
More precisely, the twist $\cF$ in \eqref{F} satisfies the factorized conditions;  
$\cF_{12}\cF_{13}\cF_{23}=\cF_{23}\cF_{13}\cF_{12}$ and 
$(\copro\otimes1)\cF=\cF_{13}\cF_{23}$, $ (1\otimes\copro)\cF=\cF_{13}\cF_{12}$. 
This is sufficient so that the cocycle condition \eqref{cocy} is satisfied.}. 
the twisted universal R-matrix $\cR^{(\cF)}$ defined in \eqref{twistedR} 
is also a solution of the Yang-Baxter equation (\ref{YBE})\,. 

\medskip 

To see the relation to our argument, let us consider the fundamental representation 
$\rho : \cU(\alg{sl}(2))\to {\rm End}(\mathbb{C}^2)$\,, 
\begin{align}
\rho(\gen{e})= e_{12} =\biggl( \; \begin{matrix} 0 & ~~1 \\ 0 & ~~0 \end{matrix}\; \biggr)\,, 
\quad 
\rho(\gen{h})=e_{11}-e_{22} =\biggl( \; \begin{matrix} 1 & 0 \\ 0 & -1\end{matrix}\; \biggr)\,,
\quad 
\rho(\gen{f})= e_{21} =\biggl( \; \begin{matrix} 0 & ~~0 \\ 1 & ~~0\end{matrix}\; \biggr)\,, \notag
\end{align} 
where $e_{ij}$ is a $2\times 2$ matrix of unity $(e_{ij})_{kl}=\de_{ik}\de_{jl}$. 
This is naturally lifted to the Yangian evaluation representation 
$\rho_u : \cY(\alg{sl}(2))\to {\rm End}(\mathbb{C}^2)$
with a spectral parameter $u\in \mathbb{C}$\,.  

\medskip 

For the Yang-Baxter equation with twisted R-matrix, one may consider 
the image of both sides by $(\rho_u\otimes\rho_v\otimes1)$\,,  
\begin{align}\label{RTT}
R_{12}^{(\cF)}(u-v)T_{1}^{(\cF)}(u)T_{2}^{(\cF)}(v)=T_{2}^{(\cF)}(v)T_{1}^{(\cF)}(u)R_{12}^{(\cF)}(u-v)\,.   
\end{align} 
Here $T_i^{(\cF)}:=(\rho_u\otimes\rho_v\otimes1)\cR_{i3}^{(\cF)}$ with $i=1,2$. 
This is the RTT relation of twisted Yangian $\cY_\xi(\alg{sl}(2))$\footnote{
This algebra is often called the deformed Yangian in some literatures. On the other hand, 
the word, twisted Yangian, is sometimes utilized to refer to the remnant of Yangian at boundary.} 
\cite{KST}. The fundamental R-matrix $R^{(\cF)}_{12}(u)\in {\rm End}(\mathbb{C}^2\otimes \mathbb{C}^2)$
and T-operator $T^{(\cF)}(u)\in  {\rm End}(\mathbb{C}^2)\otimes \cY_\xi(\alg{sl}(2))[[u^{-1}]]$
are computed as 
\begin{align}
R^{(\cF)}_{12}(u) &=(\rho\otimes\rho)\cF_{21}( 1-P_{12}u^{-1} ) \cF^{-1}   
=(\rho\otimes\rho)\cF_{21}\cF^{-1} -P_{12} u^{-1}  \no \\
&=1+\xi e_{21}\otimes (e_{11}-e_{22})-\xi (e_{11}-e_{22})\otimes e_{21} +\xi^2 e_{21}\otimes e_{21} 
- e_{ij}\otimes e_{ji} u^{-1}\,,  
\label{fundR} \\
T^{(\cF)}(u) &= (\rho\otimes1) \cF_{21} \cdot T(u) \cdot (\rho\otimes1) \cF^{-1}
=\biggl( \; \begin{matrix} 1 & 0 \\ \xi \gen{h} & 1 \end{matrix}\; \biggr) T(u) 
\biggl( \; \begin{matrix} \ga^{1/2} & 0 \\ 0 & \ga^{-1/2} \end{matrix}\; \biggr)\,, 
\end{align}
where $T(u)$ is the generating function of the undeformed Yangian 
$T(u)\in  {\rm End}(\mathbb{C}^2)\otimes \cY(\alg{sl}(2))[[u^{-1}]]$
and $\ga^{\pm1/2}$ are interpreted as the formal power-series, 
\[
\ga^{\pm1/2}=(1-2\xi \gen{f})^{\pm1/2}
=1+\sum_{n=1}^\infty\frac{\xi^n}{n!}\prod_{l=0}^{n-1}(2l\mp1)\gen{f}^n\,.
\]  
The expression of $(\rho\otimes\rho)\cF_{21}\cF^{-1}$  in \eqref{fundR} 
was firstly described by Zakrzewski \cite{Zak}. The universal form of it was 
obtained in \cite{BH}. 
The related quantum group is a non-standard quantization 
of $\cU(\alg{sl}(2))$ given by \cite{Ohn}, 
which is a $q$-deformed Poincar\'e algebra we have already encountered.

\subsection{The Jordanian pull-back \label{sec:pullback} } 

We are now ready to compare the classical integrable structure of \sch sigma models 
to quantum Jordanian twists. 

\medskip 

In general, the T-operators and R-matrix in \eqref{RTT} in quantum theory correspond  
to the monodromy matrices and the classical $r$-matrix in classical theory. 
In the present case, the untwisted T-operator $T(u)$ corresponds to  
the gauge transformed monodromy matrices $[U^{R_{\pm}}(\la_{R_{\pm}})]^{\cF}$ associated to 
the rational Lax pair $\cL^{R_\pm}_\mu(x;\la_{R_\pm})$ in \eqref{rflatlax}. On the other hand, 
the classical analogue of the twisted T-operator $T^{(\cF)}(u)$  
corresponds to the monodromy matrices $U^{R_{\pm}}(\la_{R_{\pm}})$ obtained from the 
anisotropic right Lax pairs $\cA^{R_\pm}_\mu(x;\la_{R_\pm})$ in \eqref{right lax}. 
The relation of the quantities is depicted in Fig.\,\ref{fig:twist}. 

\medskip 

Indeed, the non-local gauge transformation $\cF(x)$ in \eqref{twist} is parallel to 
the classical analogue of the quantum Jordanian twist $\cF$ in \eqref{F}, 
but the direction is opposite\footnote{%
In this section, we only discuss $\cF(x):=\cF^+(x)$ and the argument for $\cF^-(x)$
is completely parallel.}. 
The gauge transformation corresponds to 
the inverse of the quantum Jordanian twist (see Fig.\,\ref{fig:twist}). 
Thus the gauge transformation should be called the classical Jordanian twist. 
The point is that the anisotropic Lax pairs $\cA^{R_\pm}_\mu(x;\la_{R_\pm})$  
are already Jordanian twisted at the classical level. The twist can be undone by performing 
the gauge transformation and hence the isotropic Lax pairs $\cL^{R_\pm}_\mu(x;\la_{R_\pm})$ 
can be obtained. This is the mathematical background behind the exotic symmetry. 
That is, the exotic symmetry is identified with a twisted Yangian.

\begin{figure}[tbp]
\begin{align}
\begin{CD}
\text{Quantum}\quad  @. \cY(\alg{sl}(2)) \ni T(u) @> \cF >\text{Jordanian twist}> 
T^{(\cF)}(u) \in \cY_\xi (\alg{sl}(2))\text{\cite{KST}}  \\
\text{v.s.} @.  @ |  @| \\
\text{Classical}\quad @. \cY_{\rm cl}(\alg{sl}(2)) 
\ni [U^{R}(\la_{R})]^{\cF}  @ <\cF (x) <\text{Non-local gauge trans.} < 
U^{R} (\la_{R})\in \text{Exotic symm. \cite{exotic} }  
\end{CD} \no 
\end{align}
\caption{\footnotesize The correspondence between quantum and classical twists.  
The twisted T-operator (upper right) is the quantum analogue of 
the anisotropic right monodromy (bottom right) on one hand. 
The gauge transformed isotropic monodromy (bottom left) is corresponding to 
the undeformed Yangian T-operator (upper left) on the other hand. 
The classical Jordanian twist (bottom middle) is working in the opposite 
direction in contrast to the quantum one (upper middle).
\label{fig:twist}}
\end{figure}

\medskip 

Let us compare the quantum T-operator $T(u)$ to the monodromy matrix   
$[U^{R}(\la_{R})]^{\cF}$, where we take either of $\lambda_{R_{\pm}}$ 
and write it as $\lambda_R$\,. Then $T(u)$ and $[U^{R}(\la_{R})]^{\cF}$ are given by
\begin{align} 
\label{T}
T(u)&=\cF_{21}^{-1} T^{(\cF)}(u) \cF 
=\biggl( \; \begin{matrix} 1 & 0 \\ -\xi \gen{h} & ~1 \end{matrix}\; \biggr) ~T^{(\cF)}(u)\, 
\biggl( \; \begin{matrix} \ga^{-1/2} & 0 \\ 0 & ~\ga^{1/2} \end{matrix}\; \biggr)\,, \\
\label{U}
[U^R(\la_R)]^{\cF} & = \cF(+\infty)^{-1}  U^R(\la_R) \cF(-\infty)  
= K^{-1} \biggl( \; \begin{matrix} {\rm e}^{\xi Q^{R,-}} & 0 \\ 
-\sqrt{2}\xi Q^{R,2} & ~{\rm e}^{-\xi Q^{R, -}} \end{matrix}\; \biggr)\,  
U^R(\la_R) K\,. 
\end{align}
Thus a canonical choice of the constant $K^\pm$ in 
\eqref{twist} turns out to be \eqref{Kfix}. 
The correspondence between \eqref{T} and \eqref{U} also tells us that 
the quantum twists $\cF_{21}$ and $\cF$ may be interpreted as 
the world-sheet {\it boundary terms} of the non-local gauge 
transformations $\cF(+\infty)$ and $\cF(-\infty)$\,, respectively.

\section{The left description revisited \label{sec:left}} 

It is possible to describe the classical dynamics with the left description 
based on $SL(2,{\mathbb R})_{\rm L}$\,, 
though we have worked in the right description so far. 
Let us first recall the left description and then 
argue the Jordanian twist from the viewpoint of 
the left description through the left-right duality. 
In particular, the conserved currents associated with the two Yangians 
based on $SL(2,{\mathbb R})_{\rm L}$
can be rewritten in a simple form by using the twists $\cF^{\pm}$\,. 

\medskip 

The Lax pairs in the left description based on $SL(2,{\mathbb R})_{\rm L}$ 
are given by \cite{BFP,KY} 
\begin{align}
\label{leftlax}
\cL^{L_\pm}_\mu(x;\la_{L_\pm}) = \frac{
\cJ^{L_\pm}_\mu -\la_{L_\pm}\ep_{\mu\nu}\cJ^{L_\pm,\nu} }{1-\la_{L_\pm}^2}
\end{align}
with the improved $SL(2,{\mathbb R})_{\rm L}$ Noether current\footnote{%
We have changed the notations from the previous work \cite{exotic}\,, 
precisely $\cJ^{L_\pm}_\mu{}_{\text{[ours]}}=j^{L_\pm}_\mu{}_{\!\text{\cite{exotic}}}$ and 
$\cL^{L_\pm}_\mu(x;\la_{L_\pm})_{\text{[ours]}}=L^{L_\pm}_\mu(x;\la_{L_\pm}){}_{\!\text{\cite{exotic}}}$\,. 
%
}, 
\begin{align}
\cJ^{L_\pm}_\mu=\partial_\mu g \cdot g^{-1}-2C\Tr(T^- J_\mu)gT^-g^{-1}
\pm\sqrt{C}\ep_{\mu\nu}\partial^\nu(gT^-g^{-1})\,. 
\label{improved}
\end{align}
The last topological terms are required so that the current satisfies  
the flatness condition. 

\medskip 

The right Lax pairs $\cA^{R_\pm}(x;\la_{R_\pm})$ in \eqref{right lax} 
are related to the left isotropic Lax pairs in \eqref{leftlax} 
through a {\it local} gauge transformation by a group element $g(x)$\,, 
\begin{align}
[\cL^{L_\pm}_\mu(x;\la_{L_\pm})]^g=\cA^{R_\pm}_\mu(x;\la_{R_\pm})\,, 
\end{align} 
under the parameter relations $\la_{L_\pm}=1/\la_{R_\pm}$\,.  
Then, taking account of the classical Jordanian twists \eqref{rflatlax}  
and the composite rule of the gauge transformations \eqref{composit}\,, 
one can find the sequence of the gauge transformations, 
\begin{align}
\bigl[\cL^{L_\pm}_\mu(x;\la_{L_\pm}) \bigr]^{g\cF^\pm}
=\bigl[\cA^{R_\pm}_\mu(x;\la_{R_\pm}) \bigr]^{\cF^\pm}
=\cL^{R_\pm}_\mu(x;\la_{R_\pm})\,. 
\label{relation}
\end{align}
The sequence of the dualities is illustrated in Fig.\,\ref{fig:duality}\,.

\begin{figure}[tbp]
\begin{align}
\begin{CD}
\cL^{L_\pm}_\mu(x;\la_{L_\pm}) @> g(x) >\text{local gauge trans. \cite{exotic}} > \cA^{R_\pm}_\mu(x;\la_{R_\pm})  \\ 
@ V g \cF^{\pm}g^{-1}(x) =\cG^\pm(x) VV 
@VV \cF^{\pm}(x) V \\
\cA^{L_\pm}_\mu(x;\la_{L_\pm}) @> g(x) >\text{local gauge trans. } > 
\cL^{R_\pm}_\mu(x;\la_{R_\pm}) 
\end{CD} \no 
\end{align}
\caption{\footnotesize The commutative diagram of the gauge transformations. 
The isotropic left/right Lax pairs $\cL^{L/R_\pm}_\mu(x;\lambda_{L/R_\pm})$ 
(upper left/bottom right) and 
the left/right Lax pairs $\cA^{L/R_\pm}_\mu(x;\lambda_{L/R_\pm})$ 
(bottom left/upper right) are related each other through the 
(non-)local gauge transformations (arrows), 
satisfying the commutative condition $g\cF^\pm=\cG^\pm g$. 
The sequence (upper left $\to$ upper right $\to$ bottom right) is given in \eqref{relation}.}
\label{fig:duality}
\end{figure}

\medskip 

Now it is obvious from \eqref{relation} that the left isotropic Lax pairs 
$\cL^{L_\pm}_\mu(x;\la_{L_\pm})$ are directly related to the right ones 
$\cL^{R_\pm}_\mu(x;\la_{R_\pm})$ through the gauge transformations 
by $g\cF^\pm(x)$\,.
This implies that the left flat currents \eqref{improved} should be written as 
a conjugation of the right flat current \eqref{rflatcur}\,, namely  
\begin{align}
\cJ^{L_\pm}_\mu(x)=\partial_\mu(g\cF^\pm(x))\cdot(g \cF^\pm(x))^{-1}~. 
\label{leftsimple}
\end{align}
Indeed, the simple expressions in (\ref{leftsimple}) agree with the currents in \eqref{improved}. 
It is now manifest that $\cJ^{L_\pm}_\mu$ satisfy the flatness conditions. 
Note that the currents in \eqref{leftsimple} are obtained from 
the $SL(2,{\mathbb R})_{\rm L}$ current $\partial_{\mu}g\cdot g^{-1}$ in 
$SL(2,\mathbb{R})$ principal chiral models by the replacement
\begin{eqnarray}
g \to g^\pm:=g\cF^{\pm}\,. 
\end{eqnarray}
The dictionary of the left-right duality on some quantities is summarized below:
\begin{align}
\text{Current:}
&& (g^\pm)^{-1}\cJ^{L_\pm}_\mu g^\pm&=-\cJ^{R_\pm}_\mu\,, \no \\
\text{Lax pair:} && (g^\pm)^{-1}\bigl(\partial_\mu-\cL^{L_\pm}_\mu(x;\la_{L_\pm})\bigr)g^\pm 
&=\partial_\mu-\cL^{R_\pm}_\mu(x;\la_{R_\pm})\,, \no \\
\text{Spectral parameter:} &&\la_{L_\pm} &= 1/\la_{R_\pm}\,, \no \\ 
\text{Monodromy matrix:} &&g^\pm(+\infty)^{-1}\cU^{L_\pm}(\la_{L_\pm})g^\pm(-\infty)
&=\bigl[\cU^{R\pm}(\la_{R_\pm})\bigr]^{\cF^\pm}\,. \no
\end{align}
Note that the first relations for the currents are not the gauge transformations but identities. 
The currents are transformed in the same way as the Lax pairs like in (\ref{gtrf}).  
The others are associated with the gauge transformations by $g^{\pm}$\,. 

\medskip 

Before closing this section, let us comment on other expressions of Lax pairs in the left description, 
$\cA^{L_\pm}_\mu(x;\lambda_{L_\pm})$\,, which appears in Fig.\ref{fig:duality}\,. 
It is easy to derive the expressions, 
\begin{align}
& \cA^{L_\pm}_\mu(x;\la_{L_\pm}) =[\cL^{R_{\pm}}_\mu(x;\la_{R_\pm})]^{g^{-1}} \label{la} \\
&= \frac{1}{1-\la_{L_\pm}^2} \Bigl[\partial_\mu g g^{-1} 
 -\la_{L_\pm} \ep_{\mu\nu}(\partial^\nu g g^{-1} + \cG^{\pm}(x)^{-1}\partial^{\nu} \cG^{\pm}(x)  ) 
 +\la_{L_\pm}^2 \cG^{\pm}(x)^{-1}\partial^\nu \cG^{\pm}(x)  \Bigr]\,, \no
\end{align}
where we have introduced the left non-local field $\cG^\pm$ so that the diagram in Fig.\,\ref{fig:duality} is commutative;
\begin{align}
\cG^{\pm}(x):= g\cF^{\pm} g^{-1}(x)\,.  
\end{align}
On the other hand, $\cA^{R_\pm}_\mu(x;\la_{R_\pm})$ in (\ref{right lax}) can be rewritten as 
\begin{align}
& \cA^{R_\pm}_\mu(x;\la_{R_\pm})=[\cL^{L_\pm}_\mu(x;\la_{L_\pm})]^{g} \label{ra} \\
&= \frac{1}{1-\la_{R_\pm}^2} \Bigl[-g^{-1}\partial_\mu g -
 \la_{R_\pm} \ep_{\mu\nu}( -g^{-1}\partial^\nu g - \partial^\nu \cF^\pm(x) \cF^\pm(x)^{-1} ) 
 -\la_{R_\pm}^2  \partial_\mu \cF^\pm(x) \cF^\pm(x)^{-1}  \Bigr]\,. \no 
\end{align}
The expressions in (\ref{la}) are quite similar to those in (\ref{ra}). 
According to this similarity, one may expect the existence of the exotic symmetry 
even in the left description. It would be interesting to construct the corresponding generators concretely.

\section{The geometric interpretation of twists \label{sec:dipole} }

The isotropic Lax pairs for \sch sigma models are obtained from the ones 
for $SL(2;\mathbb{R})$ principal chiral models by the formal replacement of the group element $g\to g\cF^\pm$, 
as we have seen in the previous section. Here we would like to consider the following questions. 
What happens to the target space by the replacement? 
How the Jordanian twists act on the Poincar\'e coordinates of the undeformed AdS$_3$?

\medskip 

As in the case of squashed S$^3$ \cite{KMY-monodromy}, 
the classical Lagrangian of \sch sigma models \eqref{action} 
is written in a {\it dipole-like} form,   
\begin{align}\label{diaction}
L=-\eta^{\mu\nu} \Tr\bigl(\cJ^{L_+}_\mu \cJ^{L_-}_\nu\bigl)
\end{align}
with the left flat currents \eqref{leftsimple}. 
It should be emphasized that both $\cJ^{L_+}$ and $\cJ^{L_-}$ are necessary 
to express the action, while the Lax pairs are constructed from either of them.   

\medskip 

Thus, in order to construct \sch spacetimes from AdS$_3$\,, it is necessary to use 
two kinds of the Jordanian twists $\cF^\pm$\,. 
The twists $g\cF^\pm$ take the values in $SL(2;\mathbb{R})$ and hence 
the new angle variables $(v^\pm,\rho^\pm,u^\pm)$ are formally introduced 
via the following relations,  
\begin{align} \label{dicoord}
g={\rm e}^{2v T^+}{\rm e}^{2\rho T^2}{\rm e}^{2u T^-}  \quad \to \quad 
g^\pm= g\cF^\pm =:{\rm e}^{2v^\pm T^+}{\rm e}^{2\rho^\pm T^2}{\rm e}^{2u^\pm T^-} ~.  
\end{align}
We refer to the above coordinates as the {\it dipole coordinates}. 

\medskip 

Plugging the definitions of $\cF^\pm$ in \eqref{Fpm} into \eqref{dicoord}, 
the dipole coordinates are related to the usual coordinates of \sch spacetimes, 
\begin{align}
u^\pm &=\Bigl[u \pm 2\xi 
\int^x_{-\infty}\!\! dy \, \bigl(\dot{\rho} -2{\rm e}^{-2\rho}(u\dot v\pm \xi v')\bigr)
\exp\Bigl( \pm 4\xi \int^x_y \!\! dz\,{\rm e}^{-2\rho}\dot{v}\Bigr) \Bigr] 
\exp\Bigl(\mp 4\xi \! \int_x^\infty \!\! dy\,  {\rm e}^{-2\rho} \dot{v}\Bigr)\,, \no \\
v^\pm&=v\,,  \qquad 
\rho^\pm = \rho \pm 2\xi \! \int_x^\infty \!\! dy\,  {\rm e}^{-2\rho} \dot{v}\,.
\end{align}
Here dot and prime mean time and spatial derivatives, respectively, on the world-sheet. 
In the $\xi\to0$ limit (equivalently, $C\to 0$), the dipole coordinates $(v^\pm,\rho^\pm,u^\pm)$
reduce to the Poincar\'e coordinates $(v,\rho,u)$ of AdS$_3$\,. 

\medskip 

The left currents are written in the dipole coordinates as follows:  
\begin{align}
\cJ^{L,\pm}_\mu &=\partial_\mu(g\cF^\pm)\cdot(g \cF^\pm)^{-1} \no \\
&=2T^+\bigl(\partial_\mu v^\pm-2v^\pm \partial_\mu \rho^\pm +2 {\rm e}^{-2\rho^\pm}(v^\pm)^2 \partial_\mu u^\pm\bigr)\no \\
&\quad +2T^2\bigl(\partial_\mu \rho^\pm -2{\rm e}^{-2\rho^\pm}v^\pm \partial_\mu u^\pm\bigr) 
+2T^-{\rm e}^{-2\rho^\pm}\partial_\mu u^\pm\,.    
\end{align}
Then the Lagrangian \eqref{diaction} can also be expressed in terms of the dipole coordinates as 
\begin{align}
L&=-2\eta^{\mu\nu}\bigl[-2{\rm e}^{-2\rho}\partial_\mu u \partial_\nu v 
+\partial_\mu\rho \partial_\nu\rho -C^2{\rm e}^{-2\rho}\partial_\mu v \partial_\nu v\bigr] \no\\
&=-2\eta^{\mu\nu}\bigl[-({\rm e}^{-2\rho^+}\partial_\mu u^+\partial_\nu v^- 
+{\rm e}^{-2\rho^-}\partial_\mu u^-\partial_\nu v^+) +\partial_\mu\rho^+\partial_\nu\rho^-\bigr]\,. 
\label{ac}
\end{align}
This expression is very impressive because it is quite similar to the classical action of 
$SL(2,\mathbb{R})$ principal chiral models. 
In the second line of \eqref{ac}\,, the dependence on the deformation parameter $C$ 
is encoded into the dipole coordinates. The expression in the dipole coordinates 
are very curious. It might be interpreted as principal chiral models over a kind of doubled geometry 
spanned by two distinct AdS$_3$ pieces. 
It would be interesting to elaborate the geometrical meaning of the dipole coordinates.   

\medskip 

Recall that the $SL(2,\mathbb{R})_{\rm R}$ symmetry in \eqref{level0} , which is realized in a nontrivial way 
after the gauge transformation, is {\it not} an isometry of target space 
but a symmetry of \sch sigma models. This difference should be crucial. 
So, after the gauge transformation, the action \eqref{ac} looks  
very close to principal chiral models, but still different from them.

\section{Conclusion and Discussion \label{sec:concl}}

We have proceeded to study the affine extension of $q$-Poincare algebra in the Schr\"odinger sigma models. 
It has been shown that anisotropic Lax pairs are equivalent with 
isotropic Lax pairs with flat conserved currents under non-local gauge transformations. 
In this sense, the anisotropic Lax pairs are not {\it anisotropic} but {\it isotropic} in essential. 
Then a quite non-trivial realization of the undeformed Yangian symmetry 
$\mathcal{Y}(\mathfrak{sl}(2))$ has been revealed by comparing the gauge transformation 
to a quantum Jordanian twist. As a result, the exotic symmetry found in \cite{exotic} may be interpreted 
as a Jordanian twist of $\mathcal{Y}(\mathfrak{sl}(2))$\,.

\medskip

So far, we have discussed just three-dimensional Schr\"odinger spacetime (Sch$_3$), 
but it would be interesting to consider the embedding into the string-theory context. 
The present result may be directly applicable by considering a subspace like 
Sch$_3\times$S$^1$\,. It is shown in \cite{Kame} that the string sigma models correspond to 
the Jordanian deformations of $SL(2)$ spin 
chains\footnote{The Jordanian deformations of the XXX models are originally discussed in \cite{Jordanian}.}. 

\medskip 

Another interesting issue is to argue the relation between Jordanian twists and 
TsT transformations in string theory backgrounds. 
The TsT transformations may be reinterpreted as imposing twisted periodic 
boundary conditions \cite{AAF}. The twist may be regarded as the abelian version 
of the Jordanian twist  (the Reshetikhin twist) \cite{R}. In fact, the Schr\"odinger spacetime can be realized 
by performing null Melvin twists which contain light-like T-dualities, 
while space-like T-dualities are assumed in \cite{AAF}.  
Probably, it would be possible in general to show that the null Melvin twists \cite{HRR,MMT,ABM} correspond to the Jordanian twists. 
We hope that we could report on the result in this direction in the near future. 

\medskip 

Furthermore, the understanding of the Jordanian twist in the Schr\"odinger sigma models 
would be a key ingredient to consider a generalization 
from three dimensions to higher dimensions. In particular, it would enable us to study 
infinite-dimensional symmetries in higher dimensional Schr\"odinger spacetimes.

\subsection*{Acknowledgments}

We would like to thank H.~Itoyama, H.~Kanno and S.~Moriyama for useful discussions. 
The work of IK was supported by the Japan Society for the Promotion of Science (JSPS). 
TM also would like to thank A.~Molev, K.~Oshima and H.~Yamane 
for valuable discussions and comments on the mathematical aspects.

\appendix 

\section*{Appendix}

\section{Twists and currents \label{app:twistF} }

We shall derive here the explicit forms of $\mathcal{F}^{\pm}(x)$ in \eqref{Fpm} 
and the flat conserved currents in \eqref{rflatexp}. The computations for $\cF^-(x)$ 
is completely parallel with that of $\cF^+(x)$ and hence we will concentrate on $\cF^+(x)$\,. 

\medskip 

Let us start from the expression (\ref{twist}). 
The path-ordered factor can be rewritten as 
\begin{align} \label{pexp}
 & \pexp \Bigl[\int^x_{-\infty}\!\!\!dy~\cA^{R_+}_x(y;\infty)\Bigr] \notag \\ 
=& \exp\bigl[2\xi T^2\chi^-(x)\bigr]
\exp\bigl[-2\xi T^-(\chi^2(x)-\tfrac{1}{2}Q^{R,2})\bigr]
\exp\bigl[-\xi T^2Q^{R,-}\bigr]\,. 
\end{align}
Here we have used the following identity
\begin{align}
&\quad \pexp\Bigl[\int^a_b\!\!\!dx~\bigl(T^2\psi^2(x)-T^-\psi^+(x)-T^+\psi^-(x)\bigr)\Bigr] \no \\
&={\rm e}^{\phi(a)T^2}\,\pexp\Bigl[-\int^a_b\!\!\!dx~
\bigl(T^-{\rm e}^{+\phi(x)}\psi^+(x)+T^+{\rm e}^{-\phi(x)}\psi^-(x)\bigr)\Bigr]{\rm e}^{-\phi(b)T^2}\,,
\end{align} 
where $\psi^a(x)$ ($a=\pm,2$) are arbitrary scalar functions 
and the potential $\phi(x)$ is introduced through $\partial_x\phi(x)=\psi^2(x)$\,. 
Note that $\psi^-(x)=0$ in the present case. 
Thus we obtain the following form, 
\begin{align}
\cF^+(x) 
&=\pexp \Bigl[\int^x_{-\infty}\!\!\!dy~\cA^{R_+}_x(y;\infty)\Bigr]
\exp\bigl[2\xi T^2Q^{R,-}\bigr] \no\\ 
&=\exp\bigl[2\xi T^2\chi^-(x)\bigr]
\exp\bigl[-2\xi T^-(\chi^2(x)-\tfrac{1}{2}Q^{R,2})\bigr]
\exp\bigl[+\xi T^2Q^{R,-}\bigr]\,. \label{A3}
\end{align}
With the help of the relation 
\[
{\rm e}^{h T^2}T^\pm {\rm e}^{-h T^2}={\rm e}^{\pm h}T^\pm\,,
\] 
the middle exponential factor with $T^-$ can be moved to the left-most.   
After that, the expression \eqref{Fpm} has been derived. 

\medskip 

The next is to derive \eqref{rflatexp}. We start from \eqref{rflatcur}. 
By using the relation \eqref{difeq}, it is an easy task to rewrite 
$\cJ^{R_+}_\mu(x)$ as follows: 
\begin{align} 
\cJ^{R_+}_\mu(x)&= -(g\cF^+(x))^{-1}\partial_\mu (g\cF^+(x)) 
= \cF^+(x)^{-1} \left(-J_\mu -\cA^{R_+}_\mu(x;\infty) \right) \cF^+(x) \no \\
&= \cF^+(x)^{-1} \left( -T^- \e^{-2\xi \chi^-}j^{R,+}_\mu 
+T^2 \e^{-2\xi \chi^-}j^{R,2}_\mu - T^+j^{R,-}_\mu \right) \cF^+(x)\,.   
\label{FJF}
\end{align}
In the last equality, we have used the right conserved currents $j^R_\mu$\,. 
It is convenient to express the twist as $\cF^+(x)=\e^{A(x)T^- }\e^{B(x)T^2 }$ with 
\begin{align}
A(x)=-2\xi  {\rm e}^{-2\xi \chi^-(x)} \left[\chi^2(x)-\tfrac{1}{2}Q^{R,2}\right]\,, 
\quad 
B(x)=+2\xi \left[\chi^-(x)+\tfrac{1}{2}Q^{R,-}\right]\,. 
\end{align}
By using the identities  
\begin{align}
\e^{h T^-}T^2 \e^{-h T^-}=T^2+ h T^- \,, \qquad  
\e^{h T^-}T^{+}\e^{-h T^-}=T^+ + h T^2+\tfrac{1}{2}h^2 T^-\,,  \no
\end{align}
the expression \eqref{FJF} is further rewritten as 
\begin{align}
\cJ^{R_+}_\mu&= -T^- \e^{+B} \bigl[\e^{-2\xi \chi^-}(j^{R,+}_\mu +A j^{R,2}_\mu)
+\tfrac{1}{2}A^2 j^{R,-}_\mu\bigr] \no \\
&\quad +T^2  \bigl[\e^{-2\xi \chi^-} j^{R,2}_\mu + A j^{R,-}_\mu\bigr] -T^+ \e^{-B} j^{R,-}_\mu\,. 
\end{align}
Note that the non-local fields $\chi^a$ are the potential of the conserved currents 
$\ep_{\mu\nu}\partial^\nu \chi^a=-j^{R,a}_\mu$ ($a=-,2$). 
Therefore, by using the total derivative,  each $\alg{sl}(2)$-component 
can be written into the simple form \eqref{rflatexp}, 
\begin{align}
\cJ^{R_+,+}_\mu&= \e^{+B} \bigl[\e^{-2\xi \chi^-}(j^{R,+}_\mu +A j^{R,2}_\mu) 
+\tfrac{1}{2}A^2 j^{R,-}_\mu\bigr] 
=\e^{\xi Q^{R,-} }j^{R,+}_\mu +\tfrac{1}{4\xi} \ep_{\mu\nu}\partial^\nu\bigl(\e^B A^2\bigr)\,, \no\\
\cJ^{R_+,2}_\mu &=\e^{-2\xi \chi^-} j^{R,2}_\mu + A j^{R,-}_\mu
=\tfrac{1}{2\xi} \ep_{\mu\nu}\partial^\nu A\,, \quad 
\cJ^{R_+,-}_\mu = \e^{-B} j^{R,-}_\mu = \tfrac{1}{2\xi} \ep_{\mu\nu}\partial^\nu {\rm e}^{-B}\,.   
\end{align}

\section{The current algebra for \texorpdfstring{$\cJ^{R_\pm}_\mu$}{JR} 
\label{ca:app}}

Let us compute the current algebra for $\cJ^{R_\pm}_\mu$\,. 
We are confined to $\cJ^{R_+}_\mu$ below. The same argument is applicable to 
$\cJ^{R_-}_\mu$ with the sign flip of $\sqrt{C}=2\xi$\,. 

\medskip 

It is helpful to list the Poisson brackets between the components of $J_\mu=g^{-1}\partial_\mu g$\,, 
\begin{eqnarray}
&&\Bigl\{J^-_t(x),J^2_t(y)\Bigr\}_{\rm P}=-J^-_t(x)\delta(x-y)\,, 
\label{poisson:left inv one-form} \\
&&\Bigl\{J^-_t(x),J^+_t(y)\Bigr\}_{\rm P}=-J^2_t(x)\delta(x-y)\,, \nonumber \\
&&\Bigl\{J^2_t(x),J^+_t(y)\Bigr\}_{\rm P}=-\left(J^+_t+8\xi^2J^-_t\right)(x)\delta(x-y)\,, \nonumber \\
&&\Bigl\{J^-_t(x),J^-_x(y)\Bigr\}_{\rm P}=0\,, \nonumber \\
&&\Bigl\{J^-_t(x),J^2_x(y)\Bigr\}_{\rm P}=-J^-_x(x)\delta(x-y)\,, \nonumber \\
&&\Bigl\{J^-_t(x),J^+_x(y)\Bigr\}_{\rm P}=-J^2_x(x)\delta(x-y)-\partial_x\delta(x-y)\,, \nonumber \\
&&\Bigl\{J^2_t(x),J^-_x(y)\Bigr\}_{\rm P}=J^-_x(x)\delta(x-y)\,, \nonumber \\
&&\Bigl\{J^2_t(x),J^2_x(y)\Bigr\}_{\rm P}=\partial_x\delta(x-y)\,, \nonumber \\
&&\Bigl\{J^2_t(x),J^+_x(y)\Bigr\}_{\rm P}=-J^+_x(x)\delta(x-y)\,, \nonumber \\
&&\Bigl\{J^+_t(x),J^-_x(y)\Bigr\}_{\rm P}=J^2_x(x)\delta(x-y)-\partial_x\delta(x-y)\,, \nonumber \\
&&\Bigl\{J^+_t(x),J^2_x(y)\Bigr\}_{\rm P}=\left(J^+_x+4\xi^2J^-_x\right)(x)\delta(x-y)\,, \nonumber \\
&&\Bigl\{J^+_t(x),J^+_x(y)\Bigr\}_{\rm P}=4\xi^2J^2_x(x)\delta(x-y)+4\xi^2\partial_x\delta(x-y)\,. \nonumber 
\end{eqnarray}
Note that the Poisson brackets between the spatial components are zero. 

\medskip 

Let us first compute the current algebra for $j^R_\mu(x)$\,. 
For this purpose, we need the Poisson brackets between 
$\chi^-(x)$ and $J^a_\mu(x)$\,.  
Since $\chi^-(x)$ is written as 
\begin{eqnarray}
\chi^-(x)=\frac{1}{2}\int^\infty_{-\infty}\!\!\!dy~\epsilon(x-y)J^-_t(y)\,, 
\end{eqnarray}
the Poisson brackets are given by  
\begin{eqnarray}
\Bigl\{\chi^-(x),J^a_\mu(y)\Bigr\}_{\rm P}=\frac{1}{2}
\int^\infty_{-\infty}\!\!\!dz~\epsilon(x-z)\Bigl\{J^-_t(z),J^a_\mu(y)\Bigr\}_{\rm P}\,. 
\label{B3}
\end{eqnarray}
The components of the bracket \eqref{B3} are given by 
\begin{eqnarray}
&&\Bigl\{\chi^-(x),J^-_t(y)\Bigr\}_{\rm P}=0\,, \qquad 
\Bigl\{\chi^-(x),J^-_x(y)\Bigr\}_{\rm P}=0\,, 
\label{poisson:chi^-_left inv one-form} \\
&&\Bigl\{\chi^-(x),J^2_t(y)\Bigr\}_{\rm P}=-\frac{1}{2}\epsilon(x-y)J^-_t(y)\,, \nonumber \\
&&\Bigl\{\chi^-(x),J^+_t(y)\Bigr\}_{\rm P}=-\frac{1}{2}\epsilon(x-y)J^2_t(y)\,, \nonumber \\
&&\Bigl\{\chi^-(x),J^2_x(y)\Bigr\}_{\rm P}=-\frac{1}{2}\epsilon(x-y)J^-_x(y)\,. \nonumber \\
&&\Bigl\{\chi^-(x),J^+_x(y)\Bigr\}_{\rm P}=-\frac{1}{2}\epsilon(x-y)J^2_x(y)-\delta(x-y)\,. \nonumber
\end{eqnarray}
Note that we have a trivial Poisson bracket, 
\begin{eqnarray}
&&\Bigl\{\chi^-(x),\chi^-(y)\Bigr\}_{\rm P}=0\,.  
\label{poisson:chi^-}
\end{eqnarray}
With \eqref{poisson:left inv one-form}, \eqref{poisson:chi^-_left inv one-form} and \eqref{poisson:chi^-}, 
the current algebra for $j^R_\mu$ is computed as 
\begin{eqnarray}
&&\Bigl\{j^{R,-}_t(x),j^{R,-}_t(y)\Bigr\}_{\rm P}=0\,, \label{poisson:j^R} \\
&&\Bigl\{j^{R,-}_t(x),j^{R,2}_t(y)\Bigr\}_{\rm P}={\rm e}^{2\xi\chi^-}j^{R,-}_t(x)\delta(x-y)\,, \nonumber \\
&&\Bigl\{j^{R,-}_t(x),j^{R,+}_t(y)\Bigr\}_{\rm P}=j^{R,2}_t(x)\delta(x-y)\,, \nonumber \\
&&\Bigl\{j^{R,2}_t(x),j^{R,2}_t(y)\Bigr\}_{\rm P}=-\xi\epsilon(x-y)\!\left[j^{R,2}_t(x){\rm e}^{2\xi\chi^-}j^{R,-}_t(y)
+{\rm e}^{2\xi\chi^-}j^{R,-}_t(x)j^{R,2}_t(y)\right]\,, \nonumber \\
&&\Bigl\{j^{R,2}_t(x),j^{R,+}_t(y)\Bigr\}_{\rm P}={\rm e}^{2\xi\chi^-}j^{R,+}_t(x)\delta(x-y) \nonumber \\
&&\hspace{4cm}-\xi\epsilon(x-y)\!\left[j^{R,2}_t(x)j^{R,2}_t(y)+{\rm e}^{2\xi\chi^-}j^{R,-}_t(x)j^{R,+}_t(y)\right]\,, \nonumber \\
&&\Bigl\{j^{R,+}_t(x),j^{R,+}_t(y)\Bigr\}_{\rm P}=-\xi\epsilon(x-y)\!\left[j^{R,+}_t(x)j^{R,2}_t(y)+j^{R,2}_t(x)j^{R,+}_t(y)\right]\,, \nonumber \\
&&\Bigl\{j^{R,-}_t(x),j^{R,-}_x(y)\Bigr\}_{\rm P}=0\,, \nonumber \\
&&\Bigl\{j^{R,-}_t(x),j^{R,2}_x(y)\Bigr\}_{\rm P}={\rm e}^{2\xi\chi^-}j^{R,-}_x(x)\delta(x-y)\,, \nonumber \\
&&\Bigl\{j^{R,-}_t(x),j^{R,+}_x(y)\Bigr\}_{\rm P}=j^{R,2}_x(x)\delta(x-y)+2\xi{\rm e}^{2\xi\chi^-}j^{R,-}_t(x)\delta(x-y) \nonumber \\
&&\hspace{4cm}-{\rm e}^{2\xi\chi^-(x)}\partial_x\delta(x-y)\,, \nonumber \\
&&\Bigl\{j^{R,2}_t(x),j^{R,-}_x(y)\Bigr\}_{\rm P}=-{\rm e}^{2\xi\chi^-}j^{R,-}_x(x)\delta(x-y)\,, \nonumber \\
&&\Bigl\{j^{R,2}_t(x),j^{R,2}_x(y)\Bigr\}_{\rm P}=-4\xi{\rm e}^{4\xi\chi^-}j^{R,-}_t(x)\delta(x-y)+{\rm e}^{4\xi\chi^-(x)}\partial_x\delta(x-y) \nonumber \\
&&\hspace{4cm}-\xi\epsilon(x-y)\!\left[j^{R,2}_t(x){\rm e}^{2\xi\chi^-}j^{R,-}_x(y)+{\rm e}^{2\xi\chi^-}j^{R,-}_t(x)j^{R,2}_x(y)\right]\,, \nonumber \\
&&\Bigl\{j^{R,2}_t(x),j^{R,+}_x(y)\Bigr\}_{\rm P}={\rm e}^{2\xi\chi^-}j^{R,+}_x(x)\delta(x-y) \nonumber \\
&&\hspace{4cm}-\xi\epsilon(x-y)\!\left[j^{R,2}_t(x)j^{R,2}_x(y)+{\rm e}^{2\xi\chi^-}j^{R,-}_t(x)j^{R,+}_x(y)\right]\,, \nonumber \\
&&\Bigl\{j^{R,+}_t(x),j^{R,-}_x(y)\Bigr\}_{\rm P}=-j^{R,2}_x(x)\delta(x-y)+2\xi{\rm e}^{2\xi\chi^-}j^{R,-}_t(x)\delta(x-y) \nonumber \\
&&\hspace{4cm}-{\rm e}^{2\xi\chi^-(x)}\partial_x\delta(x-y)\,, \nonumber \\
&&\Bigl\{j^{R,+}_t(x),j^{R,2}_x(y)\Bigr\}_{\rm P}=-{\rm e}^{2\xi\chi^-}j^{R,+}_x(x)\delta(x-y) \nonumber \\
&&\hspace{4cm}-\xi\epsilon(x-y)\!\left[j^{R,+}_t(x){\rm e}^{2\xi\chi^-}j^{R,-}_x(y)+j^{R,2}_t(x)j^{R,2}_x(y)\right]\,, \nonumber \\
&&\Bigl\{j^{R,+}_t(x),j^{R,+}_x(y)\Bigr\}_{\rm P}=-\xi\epsilon(x-y)\!\left[j^{R,+}_t(x)j^{R,2}_x(y)+j^{R,2}_t(x)j^{R,+}_x(y)\right]\,, \nonumber \\
&&\Bigl\{j^{R,-}_x(x),j^{R,-}_x(y)\Bigr\}_{\rm P}=0\,,  \qquad 
\Bigl\{j^{R,-}_x(x),j^{R,2}_x(y)\Bigr\}_{\rm P}=0\,, \nonumber \\
&&\Bigl\{j^{R,-}_x(x),j^{R,+}_x(y)\Bigr\}_{\rm P}=2\xi{\rm e}^{2\xi\chi^-}j^{R,-}_x(x)\delta(x-y)\,, \nonumber \\
&&\Bigl\{j^{R,2}_x(x),j^{R,2}_x(y)\Bigr\}_{\rm P}=-\xi\epsilon(x-y)\!\left[j^{R,2}_x(x){\rm e}^{2\xi\chi^-}j^{R,-}_x(y)+{\rm e}^{2\xi\chi^-}j^{R,-}_x(x)j^{R,2}_x(y)\right]\,, \nonumber \\
&&\Bigl\{j^{R,2}_x(x),j^{R,+}_x(y)\Bigr\}_{\rm P}=-\xi\epsilon(x-y)\!\left[j^{R,2}_x(x)j^{R,2}_x(y)+{\rm e}^{2\xi\chi^-}j^{R,-}_x(x)j^{R,+}_x(y)\right]\,, \nonumber \\
&&\Bigl\{j^{R,+}_x(x),j^{R,+}_x(y)\Bigr\}_{\rm P}=-\xi\epsilon(x-y)\!\Bigl[j^{R,+}_x(x)j^{R,2}_x(y)+j^{R,2}_x(x)j^{R,+}_x(y)\Bigr]\,. \nonumber
\end{eqnarray}

\medskip

The next task is to consider the Poisson brackets between ${\rm e}^{-2\xi \chi^-(x)}$ and $j^R_\mu(x)$\,, 
which are evaluated as follows, 
\begin{eqnarray}
\Bigl\{{\rm e}^{-2\xi\chi^-(x)},j^{R,a}_\mu(y)\Bigr\}_{\rm P}
=\xi{\rm e}^{-2\xi\chi^-(x)}\int^\infty_{-\infty}\!\!\!dz~\epsilon(x-z)\Bigl\{j^{R,-}_t(z),j^{R,a}_\mu(y)\Bigr\}_{\rm P}\,. 
\label{B7}
\end{eqnarray}
The components of the bracket \eqref{B7} are given by 
\begin{eqnarray}
&&\Bigl\{{\rm e}^{-2\xi\chi^-(x)},j^{R,-}_t(y)\Bigr\}_{\rm P}=0\,, \qquad
\Bigl\{{\rm e}^{-2\xi\chi^-(x)},j^{R,-}_x(y)\Bigr\}_{\rm P}=0\,,
\label{poisson:chi^-_j^R} \\
&&\Bigl\{{\rm e}^{-2\xi\chi^-(x)},j^{R,2}_t(y)\Bigr\}_{\rm P}=\xi\epsilon(x-y){\rm e}^{-2\xi\chi^-(x)}{\rm e}^{2\xi\chi^-}j^{R,-}_t(y)\,, \nonumber \\
&&\Bigl\{{\rm e}^{-2\xi\chi^-(x)},j^{R,+}_t(y)\Bigr\}_{\rm P}=\xi\epsilon(x-y){\rm e}^{-2\xi\chi^-(x)}j^{R,2}_t(y)\,, \nonumber \\
&&\Bigl\{{\rm e}^{-2\xi\chi^-(x)},j^{R,2}_x(y)\Bigr\}_{\rm P}=\xi\epsilon(x-y){\rm e}^{-2\xi\chi^-(x)}{\rm e}^{2\xi\chi^-}j^{R,-}_x(y)\,, \nonumber \\
&&\Bigl\{{\rm e}^{-2\xi\chi^-(x)},j^{R,+}_x(y)\Bigr\}_{\rm P}=\xi\epsilon(x-y){\rm e}^{-2\xi\chi^-(x)}j^{R,2}_x(y)-2\xi\delta(x-y)\,. \nonumber
\end{eqnarray}
The brackets between $\chi^2(x)$ and $j^R_\mu(x)$ are also computed similarly, 
\begin{eqnarray}
\Bigl\{\chi^2(x),j^{R,a}_\mu(y)\Bigr\}_{\rm P}=-\frac{1}{2}\int^\infty_{-\infty}\!\!\!dz~\epsilon(x-z)\Bigl\{j^{R,2}_t(z),j^{R,a}_\mu(y)\Bigr\}_{\rm P}\,. \label{B9}
\end{eqnarray}
The components of the bracket \eqref{B9} are given by 
\begin{eqnarray}
\Bigl\{\chi^2(x),j^{R,-}_t(y)\Bigr\}_{\rm P}&=&\frac{1}{2}\epsilon(x-y){\rm e}^{2\xi\chi^-}j^{R,-}_t(y)\,, 
\label{poisson:chi^2_j^R} \\
\Bigl\{\chi^2(x),j^{R,2}_t(y)\Bigr\}_{\rm P}
&=&-\frac{1}{2}\sinh\left(\xi Q^{R,-}\right)j^{R,2}_t(y)-\frac{\xi}{2}Q^{R,2}{\rm e}^{2\xi\chi^-}j^{R,-}_t(y) \nonumber \\
&&-\frac{1}{2}\epsilon(x-y)\left({\rm e}^{2\xi\chi^-(x)}-{\rm e}^{2\xi\chi^-(y)}\right)j^{R,2}_t(y)\nonumber \\
&&-\xi\epsilon(x-y)\left(\chi^2(x)-\chi^2(y)\right){\rm e}^{2\xi\chi^-}j^{R,-}_t(y)\,, \nonumber \\
\Bigl\{\chi^2(x),j^{R,+}_t(y)\Bigr\}_{\rm P}&=&-\frac{1}{2}\sinh\left(\xi Q^{R,-}\right)j^{R,+}_t(y)-\frac{1}{2}\epsilon(x-y){\rm e}^{2\xi\chi^-(x)}j^{R,+}_t(y) \nonumber \\
&&-\frac{\xi}{2}Q^{R,2}j^{R,2}_t(y)-\xi\epsilon(x-y)\left(\chi^2(x)-\chi^2(y)\right)j^{R,2}_t(y)\,, \nonumber \\
\Bigl\{\chi^2(x),j^{R,-}_x(y)\Bigr\}_{\rm P}&=&\frac{1}{2}\epsilon(x-y){\rm e}^{2\xi\chi^-}j^{R,-}_x(y)\,, \nonumber \\
\Bigl\{\chi^2(x),j^{R,2}_x(y)\Bigr\}_{\rm P}&=&{\rm e}^{4\xi\chi^-(x)}\delta(x-y) 
-\frac{1}{2}\sinh\left(\xi Q^{R,-}\right)j^{R,2}_x(y)-\frac{\xi}{2}Q^{R,2}{\rm e}^{2\xi\chi^-}j^{R,-}_x(y) \nonumber \\
&&-\frac{1}{2}\epsilon(x-y)\left({\rm e}^{2\xi\chi^-(x)}-{\rm e}^{2\xi\chi^-(y)}\right)j^{R,2}_x(y)\nonumber \\
&&-\xi\epsilon(x-y)\left(\chi^2(x)-\chi^2(y)\right){\rm e}^{2\xi\chi^-}j^{R,-}_x(y)\,, \nonumber \\
\Bigl\{\chi^2(x),j^{R,+}_x(y)\Bigr\}_{\rm P}&=&-\frac{1}{2}\sinh\left(\xi Q^{R,-}\right)j^{R,+}_x(y)-\frac{1}{2}\epsilon(x-y){\rm e}^{2\xi\chi^-(x)}j^{R,+}_x(y) \nonumber \\
&&-\frac{\xi}{2}Q^{R,2}j^{R,2}_x(y)-\xi\epsilon(x-y)\left(\chi^2(x)-\chi^2(y)\right)j^{R,2}_x(y)\,. \nonumber
\end{eqnarray}

\medskip 

Then we need to compute the Poisson bracket including $\chi^-(x)$ and $\chi^2(y)$ such as  
\begin{eqnarray}
\Bigl\{{\rm e}^{-2\xi\chi^-(x)},\chi^2(y)\Bigr\}_{\rm P} 
&=&-\frac{1}{2}\sinh\left(\xi Q^{R,-}\right){\rm e}^{-2\xi\chi^-(x)} \nonumber \\
&&-\frac{1}{2}\epsilon(x-y)\left[1-{\rm e}^{-2\xi\chi^-(x)}{\rm e}^{2\xi\chi^-(y)}\right]\,. 
\label{poisson:chi^-_chi^2}
\end{eqnarray}
Similarly, the Poisson bracket between $\chi^2(x)$ and $\chi^2(y)$ can be evaluated as 
\begin{eqnarray}
\Bigl\{\chi^2(x),\chi^2(y)\Bigr\}_{\rm P} 
&=&\frac{1}{2}\sinh\left(\xi Q^{R,-}\right)\left(\chi^2(x)-\chi^2(y)\right) \nonumber \\
&&+\frac{1}{4}Q^{R,2}\left({\rm e}^{2\xi\chi^-(x)}-{\rm e}^{2\xi\chi^-(y)}\right) \nonumber \\
&&+\frac{1}{2}\epsilon(x-y)\left(\chi^2(x)-\chi^2(y)\right)\left({\rm e}^{2\xi\chi^-(x)}-{\rm e}^{2\xi\chi^-(y)}\right)\,. 
\label{poisson:chi^2}
\end{eqnarray}

\medskip 

At this stage, let us introduce a flat and conserved current $\cI_\mu$ like 
\begin{eqnarray}
&&\cI^-_\mu=\epsilon_{\mu\nu}\partial^\nu\left(\frac{2}{\xi}{\rm e}^{-2\xi\chi^-}\right)\,, 
\qquad 
\cI^2_\mu=-\epsilon_{\mu\nu}\partial^\nu\left({\rm e}^{-2\xi\chi^-}\chi^2\right)\,, \label{cI}\\
&&\cI^+_\mu=j^{R,+}_\mu+\xi\epsilon_{\mu\nu}\partial^\nu\left[{\rm e}^{-2\xi\chi}\left(\chi^2\right)^2\right]\,. \nonumber
\end{eqnarray}
The flat and conserved current ${\mathcal J}^{R_+}_\mu$ is related to ${\mathcal I}_\mu$ through 
\begin{eqnarray}
&&{\mathcal J}^{R_+,-}_\mu={\rm e}^{-\xi Q^{R,-}}\cI^-_\mu\,, \qquad 
{\mathcal J}^{R_+,2}_\mu=\cI^2_\mu+\xi Q^{R,2}\cI^-_\mu\,, \\
&&{\mathcal J}^{R_+,+}_\mu={\rm e}^{\xi Q^{R,-}}\left(\cI^+_\mu 
+ \xi Q^{R,2}\cI^2_\mu+\frac{\xi^2}{2}\left(Q^{R,2}\right)^2\cI^-_\mu\right)\,. \nonumber 
\end{eqnarray}
Using \eqref{poisson:j^R}\,, \eqref{poisson:chi^-_j^R}\,, \eqref{poisson:chi^2_j^R}\,, 
\eqref{poisson:chi^-_chi^2} and \eqref{poisson:chi^2}\,, 
we can compute the current algebra for ${\mathcal I}_{\mu}$\,. 
The Poisson brackets of the time components are given by 
\begin{eqnarray}
\Bigl\{\cI^-_t(x),\cI^-_t(y)\Bigr\}_{\rm P}&=&0\,, \label{I_t-I_t Poisson} \\
\Bigl\{\cI^-_t(x),\cI^2_t(y)\Bigr\}_{\rm P}&=&\cI^-_t(x)\delta(x-y)
+\xi^2{\mathcal Q}^-_{(0)}\cI^-_t(x)\cI^-_t(y)\,, \nonumber \\
\Bigl\{\cI^-_t(x),\cI^+_t(y)\Bigr\}_{\rm P}&=&\cI^2_t(x)\delta(x-y)
+\xi^2{\mathcal Q}^-_{(0)}\cI^-_t(x)\cI^2_t(y)\,, \nonumber \\
\Bigl\{\cI^2_t(x),\cI^2_t(y)\Bigr\}_{\rm P}&=&0\,, \nonumber \\
\Bigl\{\cI^2_t(x),\cI^+_t(y)\Bigr\}_{\rm P}&=&\cI^+_t(x)\delta(x-y)
+\xi^2{\mathcal Q}^-_{(0)}\cI^-_t(x)\cI^+_t(y)\,, \nonumber \\
\Bigl\{\cI^+_t(x),\cI^+_t(y)\Bigr\}_{\rm P}&=&
\xi^2{\mathcal Q}^-_{(0)}\Bigl[\cI^2_t(x)\cI^+_t(y)-\cI^+_t(x)\cI^2_t(y)\Bigr]\,. 
\nonumber 
\end{eqnarray}
The Poisson brackets of the time and spatial components are given by
\begin{eqnarray}
\Bigl\{\cI^-_t(x),\cI^-_x(y)\Bigr\}_{\rm P}&=&0\,, \label{I_t-I_x Poisson} \\
\Bigl\{\cI^-_t(x),\cI^2_x(y)\Bigr\}_{\rm P}&=&\cI^-_x(x)\delta(x-y)
+\xi^2{\mathcal Q}^-_{(0)}\cI^-_t(x)\cI^-_x(y)\,, \nonumber \\
\Bigl\{\cI^-_t(x),\cI^+_x(y)\Bigr\}_{\rm P}&=&\cI^2_x(x)\delta(x-y)-\partial_x\delta(x-y)
+\xi^2{\mathcal Q}^-_{(0)}\cI^-_t(x)\cI^2_x(y)\,, \nonumber \\
\Bigl\{\cI^2_t(x),\cI^-_x(y)\Bigr\}_{\rm P}&=&-\cI^-_x(x)\delta(x-y)
-\xi^2{\mathcal Q}^-_{(0)}\cI^-_t(x)\cI^-_x(y)\,, \nonumber \\
\Bigl\{\cI^2_t(x),\cI^2_x(y)\Bigr\}_{\rm P}&=&\partial_x\delta(x-y)\,, \nonumber \\
\Bigl\{\cI^2_t(x),\cI^+_x(y)\Bigr\}_{\rm P}&=&\cI^+_x(x)\delta(x-y)
+\xi^2{\mathcal Q}^-_{(0)}\cI^-_t(x)\cI^+_x(y)\,, \nonumber \\
\Bigl\{\cI^+_t(x),\cI^-_x(y)\Bigr\}_{\rm P}&=&-\cI^2_x(x)\delta(x-y)-\partial_x\delta(x-y)
-\xi^2{\mathcal Q}^-_{(0)}\cI^2_t(x)\cI^-_x(y)\,, \nonumber \\
\Bigl\{\cI^+_t(x),\cI^2_x(y)\Bigr\}_{\rm P}&=&-\cI^+_x(x)\delta(x-y)
-\xi^2{\mathcal Q}^-_{(0)}\cI^+_t(x)\cI^-_x(y)\,, \nonumber \\
\Bigl\{\cI^+_t(x),\cI^+_x(y)\Bigr\}_{\rm P}&=&
\xi^2{\mathcal Q}^-_{(0)}\Bigl[\cI^2_t(x)\cI^+_x(y)-\cI^+_t(x)\cI^2_x(y)\Bigr]\,. 
\nonumber 
\end{eqnarray}
The Poisson brackets of the spatial components are given by
\begin{eqnarray}
\Bigl\{\cI^-_x(x),\cI^-_x(y)\Bigr\}_{\rm P}&=&0\,, \label{I_x-I_x Poisson} \\
\Bigl\{\cI^-_x(x),\cI^2_x(y)\Bigr\}_{\rm P}&=&
\xi^2{\mathcal Q}^-_{(0)}\cI^-_x(x)\cI^-_x(y)\,, \nonumber \\
\Bigl\{\cI^-_x(x),\cI^+_x(y)\Bigr\}_{\rm P}&=&
\xi^2{\mathcal Q}^-_{(0)}\cI^-_x(x)\cI^2_x(y)\,, \nonumber \\
\Bigl\{\cI^2_x(x),\cI^2_x(y)\Bigr\}_{\rm P}&=&0\,, \nonumber \\
\Bigl\{\cI^2_x(x),\cI^+_x(y)\Bigr\}_{\rm P}&=&
\xi^2{\mathcal Q}^-_{(0)}\cI^-_x(x)\cI^+_x(y)\,, \nonumber \\
\Bigl\{\cI^+_x(x),\cI^+_x(y)\Bigr\}_{\rm P}&=&
\xi^2{\mathcal Q}^-_{(0)}\Bigl[\cI^2_x(x)\cI^+_x(y)-\cI^+_x(x)\cI^2_x(y)\Bigr]\,. 
\nonumber 
\end{eqnarray}
Here ${\mathcal Q}_{(0)}^-$ is defined as 
\begin{eqnarray}
{\mathcal Q}_{(0)}^-:=\int^\infty_{-\infty}\!\!\!dx~\cI^-_t(x)
=\frac{1}{\xi}\sinh\left(\xi Q^{R,-}\right)\,. 
\end{eqnarray}

\medskip 

Finally let us compute the current algebra for ${\mathcal J}^{R_+}_\mu$\,. 
For this purpose, we still need the Poisson brackets between $Q^{R,a}$ $(a=-,2)$ and $\cI_\mu(x)$\,. 

\medskip 

It is first necessary to compute the Poisson brackets between $Q^{R,a}$ $(a=-,2)$ and $j^R_\mu(x)$\,. 
Those can be obtained from the Poisson brackets \eqref{poisson:chi^-_j^R} and \eqref{poisson:chi^2_j^R} 
by taking the $x\to\infty$ limit, in which $\chi^a(x)$ is replaced by $-\frac{1}{2}Q^{R,a}$\,. 
Thus the resulting brackets are given by 
\begin{eqnarray}
\Bigl\{Q^{R,-},j^{R,-}_\mu(x)\Bigr\}_{\rm P}&=&0\,, \nonumber \\
\Bigl\{Q^{R,-},j^{R,2}_\mu(x)\Bigr\}_{\rm P}&=&{\rm e}^{2\xi\chi^-}j^{R,-}_\mu(x)\,, \nonumber \\
\Bigl\{Q^{R,-},j^{R,+}_\mu(x)\Bigr\}_{\rm P}&=&j^{R,2}_\mu(x)\,, \nonumber \\
\Bigl\{Q^{R,2},j^{R,-}_\mu(x)\Bigr\}_{\rm P}&=&-{\rm e}^{2\xi\chi^-}j^{R,-}_\mu(x)\,, \nonumber \\
\Bigl\{Q^{R,2},j^{R,2}_\mu(x)\Bigr\}_{\rm P}&=&-{\rm e}^{2\xi\chi^-}j^{R,2}_\mu(x)
-2\xi{\rm e}^{2\xi\chi^-}\chi^2 j^{R,-}_\mu(x) 
+\cosh\left(\xi Q^{R,-}\right)j^{R,2}_\mu(x)\,, \nonumber \\
\Bigl\{Q^{R,2},j^{R,+}_\mu(x)\Bigr\}_{\rm P}&=&-2\xi \chi^2 j^{R,2}_\mu(x)+\cosh\left(\xi Q^{R,-}\right)j^{R,+}_\mu(x)\,. \nonumber 
\end{eqnarray}
In the same way, from \eqref{poisson:chi^-}, \eqref{poisson:chi^-_chi^2} and \eqref{poisson:chi^2}\,,  
we obtain the following Poisson brackets: 
\begin{eqnarray}
&&\Bigl\{Q^{R,-},\chi^-(x)\Bigr\}_{\rm P}=0\,, \nonumber \\
&&\Bigl\{Q^{R,-},\chi^2(x)\Bigr\}_{\rm P}=\frac{2}{\xi}\left[{\rm e}^{2\xi\chi^-(x)}-\cosh\left(\xi Q^{R,-}\right)\right]\,, \nonumber \\
&&\Bigl\{Q^{R,2},\chi^-(x)\Bigr\}_{\rm P}=\frac{2}{\xi}\left[-{\rm e}^{2\xi\chi^-(x)}+\cosh\left(\xi Q^{R,-}\right)\right]\,, \nonumber \\
&&\Bigl\{Q^{R,2},\chi^2(x)\Bigr\}_{\rm P}=\frac{1}{2}\sinh\left(\xi Q^{R,-}\right)Q^{R,2}+\cosh\left(\xi Q^{R,-}\right)\chi^2(x) 
-{\rm e}^{2\xi\chi^-(x)}\chi^2(x)\,. \quad 
\end{eqnarray}
These Poisson brackets lead to the Poisson brackets between $Q^{R,a}$ $(a=-,2)$ and ${\mathcal I}_\mu(x)$\,: 
\begin{eqnarray}
\Bigl\{Q^{R,-},\cI^-_\mu(x)\Bigr\}_{\rm P}&=&0\,, \nonumber \\
\Bigl\{Q^{R,-},\cI^2_\mu(x)\Bigr\}_{\rm P}&=&
\cosh\left(\xi Q^{R,-}\right)\cI^-_\mu(x)\,, \nonumber \\
\Bigl\{Q^{R,-},\cI^+_\mu(x)\Bigr\}_{\rm P}&=&
\cosh\left(\xi Q^{R,-}\right)\cI^2_\mu(x)\,, \nonumber \\
\Bigl\{Q^{R,2},\cI^-_\mu(x)\Bigr\}_{\rm P}&=&
-\cosh\left(\xi Q^{R,-}\right)\cI^-_\mu(x)\,, \nonumber \\
\Bigl\{Q^{R,2},\cI^2_\mu(x)\Bigr\}_{\rm P}&=&0\,, \nonumber \\
\Bigl\{Q^{R,2},\cI^+_\mu(x)\Bigr\}_{\rm P}&=&
\cosh\left(\xi Q^{R,-}\right)\cI^+_\mu(x)\,. 
\label{Poisson QJ}
\end{eqnarray}
Since the current algebra of $\cI_\mu$ is given by \eqref{I_t-I_t Poisson}\,, \eqref{I_t-I_x Poisson} and \eqref{I_x-I_x Poisson}\,, 
with the brackets in (\ref{Poisson QJ})\,, the current algebra of ${\mathcal J}^{R_+}_\mu$ is computed as 
\begin{eqnarray}
\Bigl\{{\mathcal J}^{R_+,a}_t(x),{\mathcal J}^{R_+,b}_t(y)\Bigr\}_{\rm P}
&=&\varepsilon^{ab}_{~~c}\,{\mathcal J}^{R_+,c}_t(x)\delta(x-y)\,, \nonumber \\
\Bigl\{{\mathcal J}^{R_+,a}_t(x),{\mathcal J}^{R_+,b}_x(y)\Bigr\}_{\rm P}
&=&\varepsilon^{ab}_{~~c}\,{\mathcal J}^{R_+,c}_x(x)\delta(x-y)+\gamma^{ab}\partial_x\delta(x-y)\,, \\
\Bigl\{{\mathcal J}^{R_+,a}_x(x),{\mathcal J}^{R_+,b}_x(y)\Bigr\}_{\rm P}&=&0\,. \nonumber
\end{eqnarray}
At last, after long computations with messy expressions, we have obtained quite a simple algebra. 
Remarkably speaking, this current algebra is identical with 
the one for the usual $SL(2,{\mathbb R})$ principal chiral models, 
and it does not contain $C$ as well as non-local scalar functions.

\section{A different gauge fixing  \label{app:gfix} }

The choice of constant term in \eqref{twist} affects the relations of currents algebra, 
as we have seen in Section \ref{sec:nonlocalGT}\,. One might wonder what algebra is realized 
by a different gauge fixing. It sounds a fair question. As an example, 
let us replace the constant $K^+$ in \eqref{Kfix} by $K^+ H$ with 
$H=\e^{-\xi T^-Q^{R,2}} \e^{-\xi T^2Q^{R,-}}$\,.  
Then the twist operator is modified as 
\begin{align}
\cF^+(x) \to \cF^+(x) H = \e^{2\xi T^2\chi^-(x)} \e^{-2\xi T^-\chi^2(x)}\,. 
\end{align}
Under this replacement, the currents are transformed in the adjoint way,  
\begin{align}\label{adj}
\cJ^{R_+}_\mu(x) \to  H^{-1}\cJ^{R_+}_\mu(x) H=:\cI_\mu(x)\,,
\end{align}
where $\cI_\mu(x)$ is defined by \eqref{cI}\,. 
Since $\cI_\mu$ is also flat conserved current, an infinite number of conserved charges 
${\mathcal Q}_{(n\geq 0)}$ are generated by the BIZZ construction. For example, 
the first two charges are represented by  
\begin{eqnarray}
&&{\mathcal Q}_{(0)}=\int^\infty_{-\infty}\!\!\!dx~\cI_t(x)\,, \\
&&{\mathcal Q}_{(1)}=\frac{1}{4}\int^\infty_{-\infty}\!\!\!dx\int^\infty_{-\infty}\!\!\!dy~
\epsilon(x-y)\left[\cI_t(x),\cI_t(y)\right]-\int^\infty_{-\infty}\!\!\!dx~\cI_x(x)\,. \nonumber 
\end{eqnarray}
The expression of the charges in terms of the current is identical, 
but the current algebra of $\cI_\mu$ is completely different from that of $\cJ^{R_+}_\mu$\,.  
Hence the algebra of ${\mathcal Q}_{(n)}$ would not take the usual form of the Yangian $\cY(\alg{sl}(2))$\,, 
though these would be isomorphic because the difference is just the gauge choice. 

\medskip 

By using the current algebra of $\cI_\mu$\,, the Poisson brackets among ${\mathcal Q}_{(n)}$ 
are evaluated. The Poisson brackets at the level-zero are given by 
\begin{eqnarray}
&&\left\{{\mathcal Q}^-_{(0)},{\mathcal Q}^2_{(0)}\right\}_{\rm P}={\mathcal Q}^-_{(0)}
+\xi^2{\mathcal Q}^-_{(0)}{\mathcal Q}^-_{(0)}{\mathcal Q}^-_{(0)}\,, \nonumber \\
&&\left\{{\mathcal Q}^-_{(0)},{\mathcal Q}^+_{(0)}\right\}_{\rm P}={\mathcal Q}^2_{(0)}
+\xi^2{\mathcal Q}^-_{(0)}{\mathcal Q}^-_{(0)}{\mathcal Q}^2_{(0)}\,, \nonumber \\
&&\left\{{\mathcal Q}^2_{(0)},{\mathcal Q}^+_{(0)}\right\}_{\rm P}={\mathcal Q}^+_{(0)}
+\xi^2{\mathcal Q}^-_{(0)}{\mathcal Q}^-_{(0)}{\mathcal Q}^+_{(0)}\,. 
\end{eqnarray}
Those at the level-one are 
\begin{eqnarray}
&&\left\{{\mathcal Q}^-_{(0)},{\mathcal Q}^-_{(1)}\right\}_{\rm P}
=\left\{{\mathcal Q}^2_{(0)},{\mathcal Q}^2_{(1)}\right\}_{\rm P}=0\,, \nonumber \\
&&\left\{{\mathcal Q}^-_{(0)},{\mathcal Q}^2_{(1)}\right\}_{\rm P}={\mathcal Q}^-_{(1)}
+\xi^2{\mathcal Q}^-_{(0)}{\mathcal Q}^-_{(0)}{\mathcal Q}^-_{(1)}\,, \nonumber \\
&&\left\{{\mathcal Q}^-_{(0)},{\mathcal Q}^+_{(1)}\right\}_{\rm P}={\mathcal Q}^2_{(1)}
+\xi^2{\mathcal Q}^-_{(0)}{\mathcal Q}^-_{(0)}{\mathcal Q}^2_{(1)}\,, \nonumber \\
&&\left\{{\mathcal Q}^2_{(0)},{\mathcal Q}^-_{(1)}\right\}_{\rm P}=-{\mathcal Q}^-_{(1)}
-\xi^2{\mathcal Q}^-_{(0)}{\mathcal Q}^-_{(0)}{\mathcal Q}^-_{(1)}\,, \nonumber \\ 
&&\left\{{\mathcal Q}^2_{(0)},{\mathcal Q}^+_{(1)}\right\}_{\rm P}={\mathcal Q}^+_{(1)}
+\xi^2{\mathcal Q}^-_{(0)}{\mathcal Q}^-_{(0)}{\mathcal Q}^+_{(1)}\,, \nonumber \\
&&\left\{{\mathcal Q}^+_{(0)},{\mathcal Q}^-_{(1)}\right\}_{\rm P}=-{\mathcal Q}^2_{(1)}
-\xi^2{\mathcal Q}^-_{(0)}{\mathcal Q}^2_{(0)}{\mathcal Q}^-_{(1)}\,, \nonumber \\
&&\left\{{\mathcal Q}^+_{(0)},{\mathcal Q}^2_{(1)}\right\}_{\rm P}=-{\mathcal Q}^+_{(1)}
-\xi^2{\mathcal Q}^-_{(0)}{\mathcal Q}^+_{(0)}{\mathcal Q}^-_{(1)}\,, \nonumber \\
&&\left\{{\mathcal Q}^+_{(0)},{\mathcal Q}^+_{(1)}\right\}_{\rm P}=
\xi^2{\mathcal Q}^-_{(0)}\left[{\mathcal Q}^2_{(0)}{\mathcal Q}^+_{(1)}-{\mathcal Q}^+_{(0)}{\mathcal Q}^2_{(1)}\right]\,. 
\end{eqnarray}
At the level-two, the Poisson brackets are 
\begin{eqnarray}
&&\left\{{\mathcal Q}^-_{(1)},{\mathcal Q}^2_{(1)}\right\}_{\rm P}=
{\mathcal Q}^-_{(2)}+\frac{1}{12}\left[-2{\mathcal Q}^+_{(0)}{\mathcal Q}^-_{(0)}+\left({\mathcal Q}^2_{(0)}\right)^2\right]{\mathcal Q}^-_{(0)}-{\mathcal Q}^-_{(0)}
+\xi^2{\mathcal Q}^-_{(0)}{\mathcal Q}^-_{(1)}{\mathcal Q}^-_{(1)}\,, \nonumber \\
&&\left\{{\mathcal Q}^-_{(1)},{\mathcal Q}^+_{(1)}\right\}_{\rm P}=
{\mathcal Q}^2_{(2)}+\frac{1}{12}\left[-2{\mathcal Q}^+_{(0)}{\mathcal Q}^-_{(0)}+\left({\mathcal Q}^2_{(0)}\right)^2\right]{\mathcal Q}^2_{(0)}-{\mathcal Q}^2_{(0)}
+\xi^2{\mathcal Q}^-_{(0)}{\mathcal Q}^-_{(1)}{\mathcal Q}^2_{(1)}\,, \nonumber \\
&&\left\{{\mathcal Q}^2_{(1)},{\mathcal Q}^+_{(1)}\right\}_{\rm P}=
{\mathcal Q}^+_{(2)}+\frac{1}{12}\left[-2{\mathcal Q}^+_{(0)}{\mathcal Q}^-_{(0)}+\left({\mathcal Q}^2_{(0)}\right)^2\right]{\mathcal Q}^+_{(0)}-{\mathcal Q}^+_{(0)}
+\xi^2{\mathcal Q}^-_{(0)}{\mathcal Q}^-_{(1)}{\mathcal Q}^+_{(1)}\,. 
\end{eqnarray}
Thus the resulting algebra is regarded as a one-parameter deformation of $\cY(\alg{sl}(2))$\,. 
When $C=0$\,, the defining relations of $\cY(\alg{sl}(2))$ are reproduced. 

\medskip 

It is also interesting to see the deformed Serre relations. There are five relations. \\
i) the first one,
\begin{eqnarray}
&&\left\{\left\{{\mathcal Q}^-_{(1)},{\mathcal Q}^2_{(1)}\right\}_{\rm P},{\mathcal Q}^-_{(1)}\right\}_{\rm P}
=\frac{1}{4}{\mathcal Q}^-_{(0)}\left({\mathcal Q}^2_{(1)}{\mathcal Q}^-_{(0)}-{\mathcal Q}^-_{(1)}{\mathcal Q}^2_{(0)}\right)
+\xi^2\left\{{\mathcal Q}^-_{(0)}{\mathcal Q}^-_{(1)}{\mathcal Q}^-_{(1)},{\mathcal Q}^-_{(1)}\right\}_{\rm P}\,. \nonumber
\end{eqnarray}
ii) the second one, 
\begin{eqnarray}
&&\left\{\left\{{\mathcal Q}^-_{(1)},{\mathcal Q}^+_{(1)}\right\}_{\rm P},{\mathcal Q}^-_{(1)}\right\}_{\rm P}
+\left\{\left\{{\mathcal Q}^-_{(1)},{\mathcal Q}^2_{(1)}\right\}_{\rm P},{\mathcal Q}^2_{(1)}\right\}_{\rm P} \nonumber \\
&=&\frac{1}{4}{\mathcal Q}^-_{(0)}\left({\mathcal Q}^-_{(1)}{\mathcal Q}^+_{(0)}-{\mathcal Q}^+_{(1)}{\mathcal Q}^-_{(0)}\right)
+\frac{1}{4}{\mathcal Q}^2_{(0)}\left({\mathcal Q}^-_{(1)}{\mathcal Q}^2_{(0)}-{\mathcal Q}^2_{(1)}{\mathcal Q}^-_{(0)}\right) \nonumber \\
&&+\xi^2\left\{{\mathcal Q}^-_{(0)}{\mathcal Q}^-_{(1)}{\mathcal Q}^2_{(1)},{\mathcal Q}^-_{(1)}\right\}_{\rm P}
+\xi^2\left\{{\mathcal Q}^-_{(0)}{\mathcal Q}^-_{(1)}{\mathcal Q}^-_{(1)},{\mathcal Q}^2_{(1)}\right\}_{\rm P}\,. \nonumber 
\end{eqnarray}
iii) the third one, 
\begin{eqnarray}
&&\left\{\left\{{\mathcal Q}^-_{(1)},{\mathcal Q}^+_{(1)}\right\}_{\rm P},{\mathcal Q}^2_{(1)}\right\}_{\rm P}
=\frac{1}{4}{\mathcal Q}^2_{(0)}\left({\mathcal Q}^-_{(1)}{\mathcal Q}^+_{(0)}-{\mathcal Q}^+_{(1)}{\mathcal Q}^-_{(0)}\right)
+\xi^2\left\{{\mathcal Q}^-_{(0)}{\mathcal Q}^-_{(1)}{\mathcal Q}^2_{(1)},{\mathcal Q}^2_{(1)}\right\}_{\rm P}\,. \nonumber 
\end{eqnarray}
iv) the fourth one, 
\begin{eqnarray}
&&\left\{\left\{{\mathcal Q}^-_{(1)},{\mathcal Q}^+_{(1)}\right\}_{\rm P},{\mathcal Q}^+_{(1)}\right\}_{\rm P}
+\left\{\left\{{\mathcal Q}^2_{(1)},{\mathcal Q}^+_{(1)}\right\}_{\rm P},{\mathcal Q}^2_{(1)}\right\}_{\rm P} \nonumber \\
&=&\frac{1}{4}{\mathcal Q}^+_{(0)}\left({\mathcal Q}^-_{(1)}{\mathcal Q}^+_{(0)}-{\mathcal Q}^+_{(1)}{\mathcal Q}^-_{(0)}\right)
+\frac{1}{4}{\mathcal Q}^2_{(0)}\left({\mathcal Q}^2_{(1)}{\mathcal Q}^+_{(0)}-{\mathcal Q}^+_{(1)}{\mathcal Q}^2_{(0)}\right) \nonumber \\
&&+\xi^2\left\{{\mathcal Q}^-_{(0)}{\mathcal Q}^-_{(1)}{\mathcal Q}^2_{(1)},{\mathcal Q}^+_{(1)}\right\}_{\rm P}
+\xi^2\left\{{\mathcal Q}^-_{(0)}{\mathcal Q}^-_{(1)}{\mathcal Q}^+_{(1)},{\mathcal Q}^2_{(1)}\right\}_{\rm P} \nonumber \\
&&+\xi^2{\mathcal Q}^-_{(0)}\left(\left\{{\mathcal Q}^-_{(1)},{\mathcal Q}^2_{(1)}\right\}_{\rm P}{\mathcal Q}^+_{(1)} 
-\left\{{\mathcal Q}^2_{(1)},{\mathcal Q}^+_{(1)}\right\}_{\rm P}{\mathcal Q}^-_{(1)}\right)\,. \nonumber 
\end{eqnarray}
v) the fifth one, 
\begin{eqnarray}
\left\{\left\{{\mathcal Q}^2_{(1)},{\mathcal Q}^+_{(1)}\right\}_{\rm P},{\mathcal Q}^+_{(1)}\right\}_{\rm P} 
&=&\frac{1}{4}{\mathcal Q}^+_{(0)}\left({\mathcal Q}^2_{(1)}{\mathcal Q}^+_{(0)}
-{\mathcal Q}^+_{(1)}{\mathcal Q}^2_{(0)}\right)
+\xi^2\left\{{\mathcal Q}^-_{(0)}{\mathcal Q}^-_{(1)}{\mathcal Q}^+_{(1)},{\mathcal Q}^+_{(1)}\right\}_{\rm P} \nonumber \\
&&+\xi^2{\mathcal Q}^-_{(0)}\left(\left\{{\mathcal Q}^-_{(1)},{\mathcal Q}^+_{(1)}\right\}_{\rm P}{\mathcal Q}^+_{(1)} 
-\left\{{\mathcal Q}^2_{(1)},{\mathcal Q}^+_{(1)}\right\}_{\rm P}{\mathcal Q}^2_{(1)}\right)\,. \nonumber
\end{eqnarray}
Thus all of the Serre relations are modified due to the deformation parameter $\xi$\,. 

\medskip 

Finally, it is worth showing how the deformed Yangian is embedded into 
the exotic symmetry \cite{exotic}. 
The level-zero charges are embedded into the $q$-Poincare algebra as follows:  
\begin{align}
&{\mathcal Q}^+_{(0)}=Q^{R,+}+\tfrac{\xi}{2}\sinh(\xi Q^{R,-}) (Q^{R,2})^2\,, \no \\
&{\mathcal Q}^2_{(0)}=Q^{R,2} \cosh(\xi Q^{R,-})\,, \no\\ 
&{\mathcal Q}^-_{(0)}=\frac{\sinh (\xi Q^{R,-} )}{\xi} \,. 
\end{align}
The embedding of the level-one charges is more involved. 
The level-one charges are expressed in terms of the charges of 
the $q$-Poincare generators and $\widetilde{Q}^{R,2}$ like 
\begin{align}
&{\mathcal Q}^+_{(1)}=\tfrac{1}{4\xi } \{Q^{R,+},\{Q^{R,+},\widetilde{Q}^{R,2} \}_{\rm P} \}_{\rm P} 
-\tfrac{1}{4}\sinh(\xi Q^{R,-} )\widetilde{Q}^{R,2}Q^{R,+}
 -\tfrac{\xi}{8}(Q^{R,2})^2(\wt{Q}^{R,2}-Q^{R,2})\,, \no \\
&{\mathcal Q}^2_{(1)}=\tfrac{1}{4\xi } \bigl(\{\wt{Q}^{R,2}, Q^{R,+}\}_{\rm P}
-\cosh(\xi Q^{R,-})Q^{R,+}\bigr)\,, \qquad {\mathcal Q}^-_{(1)}=\tfrac{1}{4\xi }(\wt{Q}^{R,2} - Q^{R,2})\,. 
\end{align}
Finally, $\widetilde{Q}^{R,+}$ is contained in one of the level-two charges. 
For example, it appears in ${\mathcal Q}^-_{(2)}$\,, 
\begin{align}
{\mathcal Q}^-_{(2)}&=\tfrac{1}{8\xi^2}\bigl(\wt{Q}^{R,+}
-\cosh(\xi Q^{R,-})\{\wt{Q}^{R,2},Q^{R,+}\}_{\rm P}
-\tfrac{1}{3}\sinh^2(\xi Q^{R,-})Q^{R,+} \bigr) \no \\
&\quad +\tfrac{1}{8\xi}Q^{R,2}\sinh(\xi Q^{R,-})
\bigl(Q^{R,2}+\wt{Q}^{R,2}+\tfrac{4}{3}\sinh^2(\xi Q^{R,-})Q^{R,2}\bigr)\,. 
\end{align}

\end{document}